\documentclass[a4paper]{article}

\usepackage[english]{babel}
\usepackage{graphicx}
\usepackage{xcolor}
\usepackage{amsmath}
\numberwithin{equation}{section}
\usepackage{amsfonts}
\usepackage{amssymb}
\usepackage{authblk}
\usepackage[inline]{enumitem}
\usepackage[normalem]{ulem}

\usepackage[a4paper, margin=0.85in]{geometry}
\usepackage{setspace}
\usepackage{hyperref}
\hypersetup{          
 pdfsubject={},
 pdfkeywords={},
 unicode=true,
 breaklinks=true,
 linkcolor=blue, 
 citecolor=[rgb]{1,0.0,0.0},
 pdfborder={0 0 0},
 pdfborderstyle={},
 colorlinks=true,
 bookmarksnumbered=true,
 bookmarksopen=true,
 bookmarksopenlevel=1,
}

 \usepackage{colortbl}
 \usepackage{rotating}
 \usepackage{booktabs}
 \usepackage{multirow}

\usepackage{pdfpages}

\usepackage[sort&compress,numbers]{natbib}
\begin{document}

\title{{\bf The breaking of continuous scale invariance to discrete scale invariance: a universal quantum phase transition}}

\author[1]{Omrie Ovdat\thanks{somrie@campus.technion.ac.il}}
\author[1]{Eric Akkermans\thanks{eric@physics.technion.ac.il}}

\affil[1]{Technion, Israel Institute of Technology, Haifa 3200003}

\date{}

\maketitle

\abstract{We provide a review on the physics associated with phase transitions in which continuous scale invariance is broken into discrete scale invariance. The rich features of this transition characterized by the abrupt formation of a geometric ladder of eigenstates, low energy universality without fixed points, scale anomalies and Berezinskii-Kosterlitz-Thouless scaling is described. The important role of this transition in various celebrated single and many body quantum systems is discussed along with recent experimental realizations. Particular focus is devoted to a recent realization in graphene.}


\section{Introduction}
\label{sec:1}
Continuous scale invariance (CSI) -- a common property of physical
systems -- describes the invariance of a physical quantity $f(x)$
(e.g., the mass) when changing a control parameter $x$ (e.g., the
length). This property is expressed by a simple scaling relation,
\begin{equation}
f(ax)=b\,f(x),\label{eq: scaling relation}
\end{equation}
 satisfied $\forall a>0$ and corresponding $b(a)$, whose general
solution is the power law 
\begin{equation}
f(x)=C\,x^{\gamma}    
\end{equation}
with $\gamma=\ln b/\ln a$. Other physical systems possess the weaker discrete scale invariance
(DSI) expressed by the same scaling relation (\ref{eq: scaling relation})
but now satisfied for fixed values $a,b$ and whose solution becomes
\begin{equation}
f(x)=x^{\gamma}\,G\left(\ln x/\ln a\right), \label{eq: DSI scaling}
\end{equation}
 where $G(u+1)=G(u)$. Since $G\left(u\right)$ is a periodic function,
one can expand it in Fourier series $G\left(u\right)=\sum c_{n}e^{2\pi inu}$,
thus, 
\begin{equation}
f\left(x\right)=\sum_{n=-\infty}^{\infty}c_{n}x^{\gamma+i\frac{2\pi n}{\ln a}}.
\label{eq: DSI power law expansion}
\end{equation}
 If $f\left(x\right)$ is required to obey CSI, $G\left(u\right)$
would be constraint to fulfill the relation $G\left(u\right)=G\left(u+a_0\right)\:\forall a_0\in\mathbb{R}$.
In this case, $G\left(u\right)$ can only be a constant function,
that is, $c_{n}=0$ for all $n\neq0$ eliminating all terms with complex
exponents in \eqref{eq: DSI power law expansion}. Therefore, real exponents are a signature of CSI and complex
exponents are a signature of DSI. 

In this article we describe a variety of distinct quantum systems in which a sharp transition initiates the breaking of CSI into DSI. Essential to all these cases is a DSI phase characterized by a sudden appearance of a low energy spectrum arranged in an infinite geometric series. Accordingly, each transition is associated with exponents that change from real to complex valued at the critical point. We describe the universal properties of this transition. Particularly, in the framework of the renormalization group it is shown that universality in this case is not associated with trajectories terminating at a fixed point but with periodic flow known as a limit cycle. Intrinsic to this phenomena is a special type of scale anomaly in which residual discrete scaling symmetry remains at the quantum level. 

We discuss the physical realizations of the CSI to DSI transition and present recent experimental observations which provide evidence for the existence of the critical point and for the universal low energy features of the DSI phase. We discuss the basic ingredients that underline these features and the possibility of their occurrence in other, yet to be studied systems.

\section{The Schr\"{o}dinger $1/r^{2}$ potential}
\label{sec:The 1/r^2 -potential}

A well studied example exhibiting the breaking of CSI to DSI is given by the problem of a quantum particle in an attractive inverse
square potential \cite{Case1950b,Landau1991b} described by the Hamiltonian
($\hbar=1,\,m=1/2$) 
\begin{equation}
H_{S}=p^{2}-\lambda/r^{2}.\label{eq:H_S Schrodinger equation}
\end{equation}
This system constitutes an effective description of the ``Efimov
effect'' \cite{Efimov1970c,Efimov1971b} and plays a role in various
other systems \cite{Levy-Leblond1967,Camblong:2001zt,Kaplan:2009kr,NisoliBishop2014, DeMartinoKloepferMatrasulovEtAl2014}. 

\subsection{The spectral properties of $H_{S}$} \label{subsec: The spectral properties of H_S}

The Hamiltonian $H_{S}$ has an interesting yet disturbing property
-- the power law form of the potential matches the order of the kinetic
term. As a result, the Schr\"{o}dinger equation 
\begin{equation}
H_{S}\psi=E\psi\label{eq: 1/r^2 Schrodinger}
\end{equation}
 depends on the single dimensionless parameter $\lambda$ which raises
the question of the existence of a characteristic energy to express the
eigenvalues $E_{n}$. This absence of characteristic scale implies
the invariance of $H_{S}\psi=E\psi$ under the scale transformation \cite{Jackiw1995b}
\begin{equation}
x^{i}\rightarrow ax^{i},\,E\rightarrow a^{-2}E
\end{equation}
 which indicates that if there is one negative energy bound state
$E_{n}$ then there is an unbounded continuum of bound states which
render the Hamiltonian nonphysical and mathematically not self-adjoint
\cite{Meetz1964d,GitmanTyutinVoronov2012c}. 

The eigenstates of $H_{S}$ can be solved in terms of Bessel functions
which confirm these assertions in more detail. For $E<0$ and lowest
orbital angular momentum subspace $l=0$, the most general decaying
solution is described by the radial function %
\begin{equation}
\psi\left(r\right)\approx r^{-\frac{d-2}{2}}\left(\left(kr\right)^{-\sqrt{\lambda_{c}-\lambda}}\left(a_{1}+\mathcal{O}\left(kr\right)^{2}\right)+\left(kr\right)^{\sqrt{\lambda_{c}-\lambda}}\left(a_{2}+\mathcal{O}\left(kr\right)^{2}\right)\right)\label{eq: s-wave solution of 1/r^2 potential}
\end{equation}
where $k\equiv\sqrt{-E}$, $a_{1},\,a_{2}$ are energy independent coefficients,
$d$ is the space dimension and $\lambda_{c}\equiv(d-2)^{2}/4$ \footnote{For higher angular momentum channels $\lambda_{c}$ is larger and
given by $(d-2)^{2}/4+l\left(l+d-2\right)$ }. As seen in (\ref{eq: s-wave solution of 1/r^2 potential}), for
$\lambda>\lambda_{c}-1$, $\psi_{0}\left(r\right)$ is normalizable
$\forall\, \mathrm{Re}\, \left(E\right) < 0$ which constitutes a continuum of complex
valued bound states of $H_{S}$. Thus, for $\lambda>\lambda_{c}-1$,
$H_{S}$ is no longer self-adjoint, a property that originates from
the strong singularity of the potential and is characteristic of a
general class of potentials with high order of singularity \cite{Case1950b}. 

A simple, physically instructive procedure to deal with the absence
of self-adjointness is to remove the singular $r=0$ point by introducing
a short distance cutoff $L$ and to apply a boundary condition at
$r=L$ \cite{DeMartinoKloepferMatrasulovEtAl2014,AlbeverioHoegh-KrohnWu1981a,BeaneBedaqueChildressEtAl2001a,MuellerHo2004,BraatenPhillips2004b,HammerSwingle2006c,MorozSchmidt2009}.
The most general boundary condition is the mixed condition
\begin{equation}
L\frac{\psi'\left(L\right)}{\psi\left(L\right)}=g,\label{eq:mixed condition}
\end{equation}
 $g\in\mathbb{R}$, for which there is an infinite number of choices
each describing different short range physics.

Equipped with condition (\ref{eq:mixed condition}) the operator $H_{S}$
is now a well defined self-adjoint operator on the interval $L<r<\infty$. The spectrum of
$H_{S}$ exhibits two distinct features in the low energy $kL\ll1$
regime. For $\lambda<\lambda_{c}\equiv\left(d-2\right)^{2}/4$, the
expression of $L\psi'\left(L\right)/\psi\left(L\right)$ as given
from (\ref{eq: s-wave solution of 1/r^2 potential}) is independent
of $k$ to leading order in $kr$. As a result, equation (\ref{eq:mixed condition})
does not hold for a general choice of $g$. For $\lambda>\lambda_{c}$,
the insertion of (\ref{eq: s-wave solution of 1/r^2 potential}) into
(\ref{eq:mixed condition}) leads to 
\begin{equation}
\left(kL\right)^{2i\sqrt{\lambda-\lambda_{c}}}=e^{i\gamma}\label{eq:k relation}
\end{equation}
 where $\gamma\left(g,\lambda\right)$ is a phase that can be calculated (the explicit expression of $\gamma$ is not important for the purpose of this section). The solution of (\ref{eq:k relation}) yields a set of bound states
with energies
\begin{equation}
k_{n}=k_{0} e^{-\frac{\pi n}{\sqrt{\lambda-\lambda_{c}}}} \label{eq: Efimov spectrum}
\end{equation}
where $n\in Z$, such that $k_{n}L\ll1$ and $k_{0}\equiv\frac{1}{L}e^{\frac{\gamma}{2\sqrt{\lambda-\lambda_{c}}}}$.
 Thus, for $\lambda<\lambda_{c}\equiv\left(d-2\right)^{2}/4$, the
spectrum contains no bound states close to $E=0$, however, as $\lambda$
goes above $\lambda_{c}$, an infinite series of bound states appears.
Moreover, in this ''over-critical''
regime, the states arrange in a geometric series such that 
\begin{equation}
k_{n+1}/k_{n}=e^{-\frac{\pi}{\sqrt{\lambda-\lambda_{c}}}}.
\end{equation}
The absence of any states for $\lambda<\lambda_{c}$ is a signature
of CSI while the geometric structure of (\ref{eq: Efimov spectrum})
for $\lambda>\lambda_{c}$ is a signature of DSI since $k_n$ is invariant under $\left\{ k_{n}\right\} \to\left\{ \exp\left(-\pi/\sqrt{\lambda-\lambda_{c}}\right)k_{n}\right\} $.
Accordingly, as seen in (\ref{eq: s-wave solution of 1/r^2 potential}),
the characteristic behavior of the eigenstates for $kr\ll1$ manifests
an abrupt transition from real to complex valued exponents as $\lambda$
exceeds $\lambda_{c}$. Thus, $H_{S}$ exhibits a quantum phase transition
(QPT) at $\lambda_{c}$ between a CSI phase and a DSI phase. The characteristics
of this transition are independent of the values of $L,g$ which enter
only into the overall factor $k_{0}$ in (\ref{eq: Efimov spectrum}).
The functional dependence of $k_{n}$ on $\sqrt{\lambda-\lambda_{c}}$
is characteristic of Berezinskii-Kosterlitz-Thouless (BKT) transitions
as was identified in \cite{Kaplan:2009kr,KolomeiskyStraley1992c,Jensen2011,JensenKarchSonEtAl2010}.
Finally, the breaking of CSI to DSI in the $\lambda>\lambda_{c}$
regime constitutes a special type of scale anomaly since a residual
symmetry remains even after regularization (see Table \ref{tab:QPTtable}). 

\begin{table}
\centering{}\caption{\label{tab:QPTtable} Summary of the properties associated with the transition occurring at $\lambda = \lambda_c$ for the Hamiltonian $H_S$ given in equation \eqref{eq:H_S Schrodinger equation} on the interval $L < r < \infty$.}
\medskip{}
\begin{tabular}{ccccc}
\cmidrule{1-4} 
 & $\lambda<\lambda_{c}$ & $\lambda>\lambda_{c}-1$ & $\lambda>\lambda_{c}$ & \multirow{5}{*}{\begin{turn}{-90}
Scale anomaly $\implies$
\end{turn}}\tabularnewline
\cmidrule{1-4} 
\addlinespace
\cellcolor{yellow}Formal Hamiltonian & CSI & CSI &  CSI & \tabularnewline
\addlinespace
\cellcolor{yellow}Self-adjointness & $H=H^{\dagger}$ &  $H\neq H^{\dagger}$ &  $H\neq H^{\dagger}$ & \tabularnewline
\addlinespace
\cellcolor{yellow}Regularization with $L$ & Redundant &  Essential &  Essential & \tabularnewline
\addlinespace
\cellcolor{yellow}Symmetry of eigenspace & CSI & CSI &  DSI & \tabularnewline
\cmidrule{1-4} 
\addlinespace
\multicolumn{4}{c}{Quantum Phase Transition $\implies$} & \tabularnewline
\addlinespace
\end{tabular}
\end{table}


\subsection{Physical realizations of $H_{S}$} \label{subsec: Realizations of H_S}

A well known realization of $H_{S}$ for $\lambda>\lambda_{c}$ is
the ``Efimov effect'' \cite{Efimov1970c,Efimov1971b,BraatenHammer2006}.
In $1970$, Efimov studied the quantum problem of three identical
nucleons of mass $m$ interacting through a short range ($r_{0}$)
potential. He pointed out that when the scattering length $a$ of
the two-body interaction becomes very large, $a\gg r_{0}$, there
exists a scale-free regime for the low-energy spectrum, $\hbar^{2}/ma^{2}\ll E\ll\hbar^{2}/mr_{0}^{2}$,
where the corresponding bound-states energies follow the geometric
series $E_{n}=-E_{0}e^{-2\pi n/s_{0}}$ where $s_{0}\approx1.00624$
is a dimensionless number and $E_{0}>0$ a problem-dependent energy
scale. Efimov deduced these results from an effective Schr\"{o}dinger
equation in $d=3$ with the radial ($l=0$) attractive potential $V\left(r\right)=-\lambda/r^{2}$
with $\lambda=s_{0}+1/4>\lambda_{c}$ ($\lambda_{c}=1/4$ for $d=3$).
Despite being initially controversial, Efimov physics has turned into
an active field especially in atomic and molecular physics where the
universal spectrum has been studied experimentally \cite{kraemer2006evidence,TungJimenez-GarciaJohansenEtAl2014a,PiresUlmanisHaefnerEtAl2014,pollack2009universality,GrossShotanKokkelmansEtAl2009,LompeOttensteinSerwaneEtAl2010,NakajimaHorikoshiMukaiyamaEtAl2011,KunitskiZellerVoigtsbergerEtAl2015}
and theoretically \cite{BraatenHammer2006}. The observation of the Efimov geometric spectral
ratio $e^{2\pi/s_{0}}\approx515.028$ have been recently determined
using an ultra-cold gas of caesium atoms \cite{huang2014observation}.

In addition to the Efimov effect, the inverse square potential also
describes the interaction of a point like dipole with an electron
in three dimensions. In this case, the dipole potential is considered
as an inverse square interaction with non-isotropic coupling \cite{Camblong:2001zt}.
The Klein Gordon equation for a scalar field on an Euclidean AdS $d+1$
space time can be written in the form of (\ref{eq: 1/r^2 Schrodinger}).
The over-critical regime $\lambda>\lambda_{c}$ corresponds to the
violation of the Breitenlohner-Freedman bound \cite{Kaplan:2009kr}.

\section{Massless Dirac Coulomb system } \label{sec: Dirac Coulomb system}

The inverse square Hamiltonian (\ref{eq:H_S Schrodinger equation}),
a simple system exhibiting a rich set of phenomena, inspires studying
the ingredients which lead to the aforementioned DSI and QPT and whether
they are found in other systems. One such candidate system is described
by a massless Dirac fermion in an attractive Coulomb potential \cite{MIRANSKY1980421, PereiraNilssonCastroNeto2007, ShytovKatsnelsonLevitov2007,ShytovKatsnelsonLevitov2007d} with the scale invariant Hamiltonian ($\hbar=c=1)$
\begin{equation}
H_{D}=\gamma^{0}\gamma^{j}p_{j}-\beta/r
\label{eq: Dirac Hamiltonian}
\end{equation}
 where $\beta$ specifies the strength of the electrostatic potential,
$d$ is the space dimension and $\gamma^{\mu}$ are $d+1$ matrices
satisfying the anti-commutation relation 
\begin{equation}
\left\{ \gamma^{\mu},\gamma^{\nu}\right\} =2\eta^{\mu\nu}
\end{equation}
 with $\eta^{00}=\eta^{ii}=-1$, $i=1,\ldots,d$ and $\eta^{\mu\nu}=0$
for $\mu\neq\nu$. 

Based on the previous example, it may be anticipated that, like $H_{S}$,
$H_{D}$ will exhibit a sharp spectral transition at some critical
$\beta$ in which the singularity of the potential will ruin self-adjointness.
As detailed below, the analog analysis of the Dirac equation 
\begin{equation}
H_{D}\psi=E\psi\label{eq: Dirac equation}
\end{equation}
 confirm these assertions and details a remarkable resemblance between
the low energy features of the two systems. 

\subsection{The spectral properties of $H_{D}$}

Utilizing rotational symmetry, the angular part of equation (\ref{eq: Dirac equation})
can be solved and the radial dependence of $\psi$ is given in terms
of two functions $\Psi_{1}\left(r\right),\Psi_{2}\left(r\right)$
\cite{dong2011wave} determined by the following set of equations
\begin{align}
\Psi_{2}'\left(r\right)+\frac{\left(d-1+2K\right)}{2r}\Psi_{2}\left(r\right) & =\left(E+\frac{\beta}{r}\right)\Psi_{1}\left(r\right)\nonumber \\
-\Psi_{1}'\left(r\right)-\frac{\left(d-1-2K\right)}{2r}\Psi_{1}\left(r\right) & =\left(E+\frac{\beta}{r}\right)\Psi_{2}\left(r\right)\label{eq:radial Dirac eq}
\end{align}
 where 
\begin{equation}
K\equiv\begin{cases}
\pm\left(l+\frac{d-1}{2}\right) & d>2\\
m+1/2 & d=2
\end{cases},\label{eq:def K}
\end{equation}
$l=0,1,\ldots$ and $m\in\mathbb{Z}$ are orbital angular momentum
quantum numbers. In terms of these radial functions, the scalar product
of two eigenfunctions $\psi,\tilde{\psi}$ is given by 
\begin{equation}
\int dV\,\psi^{\dagger}\tilde{\psi}=\int dr\,r^{d-1}\left(\Psi_{1}^{\ast}\left(r\right)\tilde{\Psi}_{1}\left(r\right)+\Psi_{2}^{\ast}\left(r\right)\tilde{\Psi}_{2}\left(r\right)\right).
\end{equation}

Unlike $H_{S}$ in section \ref{sec:The 1/r^2 -potential}, the spectrum
of $H_{D}$ does not contain any bound states, a property that
reflects the absence of a mass term. As a result, the spectrum is a continuum of scattering states spanning $-\infty<E<\infty$. In the absence of bound states we explore the possible occurrence of ``quasi-bound''
states. Quasi bound states are pronounced peaks in the density of
states $\rho(E)$, embedded within the continuum spectrum. These resonances describe
a scattering process in which an almost monochromatic wave packet
is significantly delayed by $V\left(r\right)$ as compared to the
same wave packet in free propagation \cite{Friedrich2013b}. 

An elegant procedure for calculating the quasi-bound spectrum \cite{Friedrich2013b}
is to allow the energy parameter to be complex valued $E\rightarrow\epsilon\equiv E_{R}-i\frac{W}{2}$
and look for solutions of (\ref{eq:radial Dirac eq}) with no outgoing
$e^{-iEr}$ plane wave component for $r\rightarrow\infty$. The lifetime
of the resonance is given by $W^{-1}.$ Consider the lowest angular
momentum subspace $K=\pm\left(d-1\right)/2$ and $E<0$, the most
general solution with no outgoing component is given by 
\begin{eqnarray}
\begin{pmatrix}\Psi_{1}\left(r\right)\\
\Psi_{2}\left(r\right)
\end{pmatrix}\approx  r^{-\frac{d-1}{2}} & \left(\left(2iEr\right)^{\sqrt{\beta_{c}^{2}-\beta^{2}}}\left(\begin{pmatrix}a_{11}\\
a_{12} \end{pmatrix}+\mathcal{O}\left(\left|E\right|r\right)\right) \right. \nonumber \\ 
 + & \left. \left(2iEr\right)^{-\sqrt{\beta_{c}^{2}-\beta^{2}}}\left(\begin{pmatrix}a_{21}\\
a_{22}
\end{pmatrix}+\mathcal{O}\left(\left|E\right|r\right)\right)\right)
\label{eq: ingoing solution}
\end{eqnarray}
 where $\beta_{c}\equiv (d-1)/2$ \footnote{For higher angular momentum channels $\beta_c$ is larger and given by $ \left| K \right| $ where $K$ is defined as in \eqref{eq:def K} } and $a$ is a $2\times2$ energy
independent coefficient matrix.

As in the case of $H_{S}$ above it is necessary at this point to
remove the singularity of the $1/r$ potential by introducing a radial
short distance cutoff $L$ and imposing a boundary condition. To identify this explicitly, consider the case where $E=i$ in (\ref{eq: ingoing solution}).
Since (\ref{eq: ingoing solution}) is (asymptotically) an ingoing $e^{iEr}$ plane
wave solution if $E\in\mathbb{R}$, it decays exponentially for $E=i$
and $r\to\infty$. If additionally $\beta^2 > \beta_{c}^{2}-1/4$,
then (\ref{eq: ingoing solution}) is a normalizable eigenfunction
with a complex valued eignvalue which renders $H_{D}$ not self-adjoint. 

The equivalent mixed boundary condition of (\ref{eq: Dirac equation})
can be written as follows \cite{Yang1987b}
\begin{equation}
h=\frac{\Psi_{2}\left(L\right)}{\Psi_{1}\left(L\right)}\label{eq:BC Dirac}
\end{equation}
 where $h\in\mathbb{R}$ is determined by the short range physics.
Equipped with this condition $H_{D}$ is now a well defined self-adjoint operator on the interval
$L<r<\infty$. The spectrum of $H_{D}$ exhibits two distinct pictures
in the low energy $\left|E\right|L\ll1$ regime. For $\beta<\beta_{c}\equiv\left(d-1\right)/2$,
the expression of $\Psi_{2}\left(L\right)/\Psi_{1}\left(L\right)$
as given from (\ref{eq: ingoing solution}) is independent of $E$
to leading order in $\left|E\right|L$. As a result, equation (\ref{eq:BC Dirac})
does hold for a general choice of $h$. For $\beta>\beta_{c}$, the
insertion of (\ref{eq: ingoing solution}) to (\ref{eq:BC Dirac})
reduces into

\begin{equation}
\left(2iEL\right)^{2i\sqrt{\beta^{2}-\beta_{c}^{2}}}=z_{0}\label{eq:E relation dirac}
\end{equation}
 where 
\begin{equation}
z_{0}\left(h,\beta\right)\equiv\frac{h\,a_{21}-a_{22}}{a_{12}-h\,a_{11}}\label{eq:z0 explicit}
\end{equation}
 is a complex valued number\footnote{Here $z_{0}$ is not a phase like in (\ref{eq:k relation}), a reflection
of the fact that the solutions for $E$ would have an imaginary component
corresponding to a finite lifetime.} (the explicit exporession for $z_0$, which can be found in \cite{OvdatMaoJiangEtAl2017}, is not important for the purpose of this section). The solution of (\ref{eq:E relation dirac}) yields a set of quasi-bound
energies at 
\begin{equation}
E_{n}=E_{0}e^{-\frac{\pi n}{\sqrt{\beta^{2}-\beta_{c}^{2}}}}\label{eq: Efimov spectrum-Dirac}
\end{equation}
where $n\in \mathbb{Z}$, such that $\left|E_{n}\right|L\ll1$ and $E_{0}\equiv\mathrm{Re}\left(\frac{1}{2iL}z_{0}^{\frac{1}{2i\sqrt{\beta^{2}-\beta_{c}^{2}}}}\right)$. It can be directly verified that $E_R = \mathrm{Re} \, E_n < 0 $ and $W = -2 \mathrm{Im} \, E_n >0$ \cite{OvdatMaoJiangEtAl2017}.

Thus, in complete analogy with the $-\lambda/r^2$ inverse squared potential described
in section \ref{sec:The 1/r^2 -potential}, for $\beta<\beta_{c}\equiv\left(d-1\right)/2$,
the spectrum contains a CSI phase with no quasi-bound states close
to $E=0$. As $\beta$ exceeds $\beta_{c}$, an infinite series of
quasi-bound states appears which arrange in a DSI geometric series
such that 
\begin{equation}
E_{n+1}/E_{n}=e^{-\frac{\pi}{\sqrt{\beta^{2}-\beta_{c}^{2}}}}.
\end{equation}
As seen explicitly in (\ref{eq: ingoing solution}), the characteristic
behavior of the eigenstates for $\left|E\right|r\ll1$ manifests an
abrupt transition from real to complex valued exponents at $\beta=\beta_{c}$.
The characteristics of this transition are independent of the values
of $L,h$ which enter only into the overall factor $E_{0}$ in (\ref{eq:E relation dirac}). Thus, under a proper trasformation between $\lambda$ and $\beta$, Table \ref{tab:QPTtable} represents a valid and consistent description of the massless Dirac Coulomb system as well.

\subsection{Distinct features associated with spin $1/2$} \label{subsec: Distinct features associated with spin}

\begin{figure}[t]
\includegraphics[width=1\textwidth]{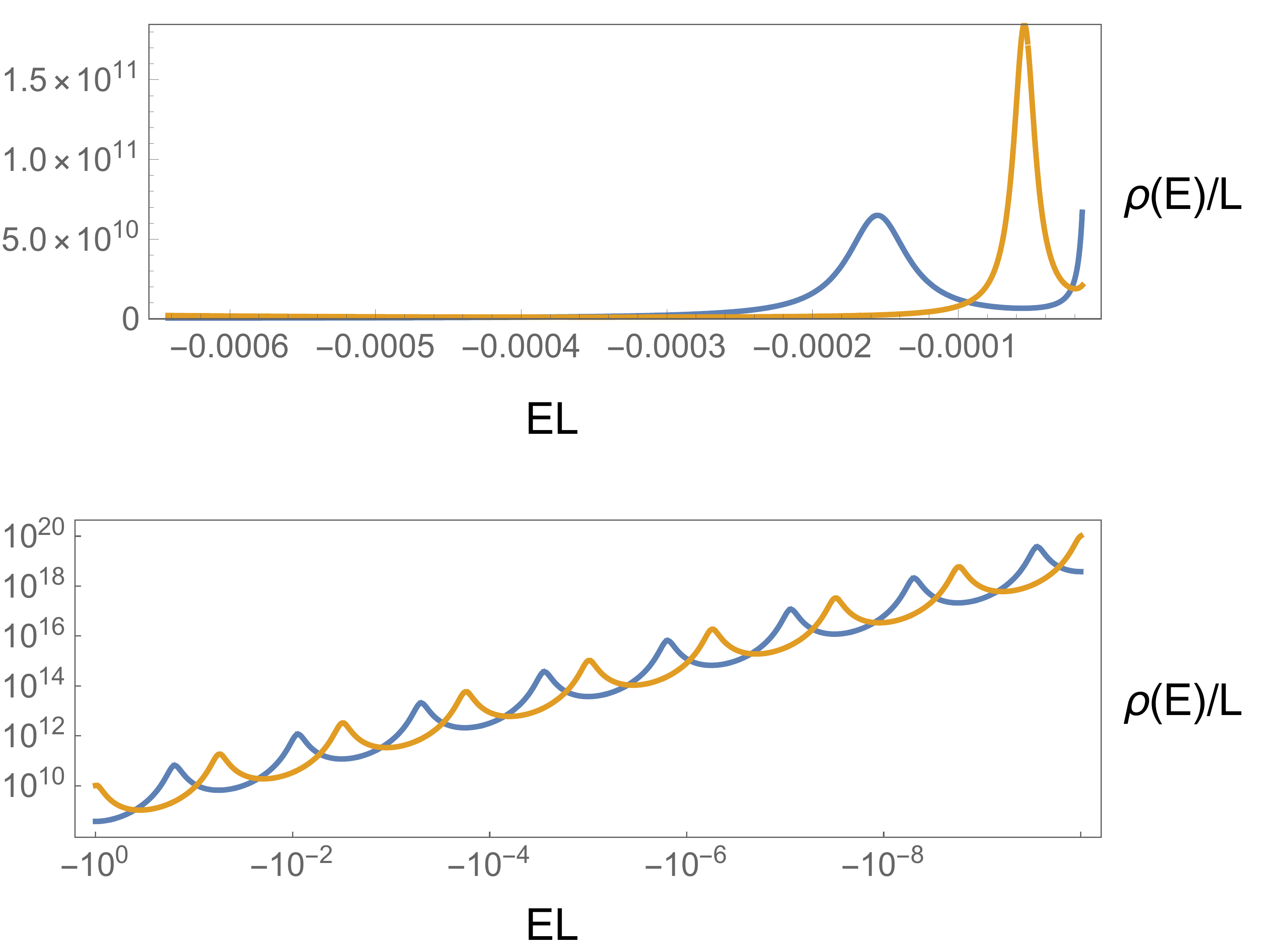}
\caption{\label{fig: Dirac efimov ladder} Density of states $\rho(E)$ of $H_D$ for $d=2$, $\beta = 1.2  > \beta_c$ and different angular momentum eigenstates. The yellow and blue curves correspond to the $m=0,-1$ angular momentum channels respectively. The boundary condition $h$ used here is the chiral boundary condition \eqref{eq: chiral BC} which breaks parity. The parameter $L$ is the short distance cutoff taken here to be $0.195 \mathrm{nm}$. The numeric values on both axis are in units of $\hbar c = 0.197 \, \mathrm{eV}\, \mu \mathrm{m} $. The set of pronounced peaks in both curves describes the quasi-bound spectrum in the overcritical regime $\beta>\beta_c$ as calculated in \eqref{eq: Efimov spectrum-Dirac}. The lower panel displays the detailed structure of the
infinite geometric ladders of the quasi-bound states in a logarithmic scale. The $m = 0,  -1$ ladders are intertwined, indicative of the breaking of parity by the boundary condition. These results are independent of the specific choice of $L$ or $h$ (provided that it breaks parity).}
\end{figure}

On top of the similarities emphasized above, an interesting difference
in the quantum phase transition exhibited by $H_{S}$ and $H_{D}$
results from the distinct spin of the associated Schr\"{o}dinger and Dirac
wave functions. Unlike the scalar Schr\"{o}dinger case, the lowest angular
momentum subspace of $H_{D}$ contains two channels corresponding
to $K=\pm (d-1)/2$. As a result, not one but two copies of geometric ladders
of the form (\ref{eq: Efimov spectrum-Dirac}) appear at $\beta=\beta_{c}$
(see Fig. \ref{fig: Dirac efimov ladder}). These two ladders may be degenerate or intertwined depending on
the choice of boundary condition in (\ref{eq:BC Dirac}). 

The breaking of the degeneracy between the ladders is directly related
to the breaking of a symmetry. To understand this point more explicitly
consider the case where $d=2$. There, in a basis where $\gamma^0 = \sigma_z$, $\gamma^1 = i\sigma_1$, $\gamma^2 = -i\sigma_2$,  $H_{D}$ is given by 
\begin{equation}
    H_D = \sigma_i p_i - \beta/r \label{eq: 2d H_D}.
\end{equation}
From \eqref{eq: 2d H_D} it is seen that $H_D$ is symmetric under the following parity transformation 
\begin{equation}
x\rightarrow-x,\,y\to y,H_{D}\to\sigma_{2}H_{D}\sigma_{2},
\end{equation}
 which in terms of $\Psi_{1}\left(r\right),\,\Psi_{2}\left(r\right)$
is equivalent to \cite{OvdatMaoJiangEtAl2017}
\begin{equation}
\Psi_{1}\left(r\right)\to\Psi_{2}\left(r\right),\,\Psi_{2}\left(r\right)\to-\Psi_{1}\left(r\right),\,m\to-m-1\label{eq:parity transformation radial}
\end{equation}
where $m$ is the orbital angular momentum. Consequently, the Dirac equation (\ref{eq:radial Dirac eq})
is invariant under (\ref{eq:parity transformation radial}), however,
the boundary condition \eqref{eq:BC Dirac} can break (\ref{eq:parity transformation radial}).
Typical choices of boundary conditions are 

\begin{enumerate}
\item \label{list:BC1} Continuously
connected constant potential $V\left(r<L\right)=-\beta/L$ \cite{PereiraKotovCastroNeto2008a}
corresponding to $h=J_{m+1}(\beta+EL)/J_{m}(\beta+EL),$ where $J_{n}(x)$
is Bessel's function. \item \label{list:BC2} Zero wavefunction of one of the spinor components \cite{ShytovKatsnelsonLevitov2007}
corresponding to $h=0$ or $h=\infty$. \item \label{list:BC3} Infinite mass
term on boundary \cite{PereiraNilssonCastroNeto2007} corresponding
to $h=1$. \item \label{list:BC4} Chiral boundary conditions  \cite{ovdat2018vacancies}
\begin{equation}
h=\begin{cases}
0 & m\geq1\\
\infty & m\leq0
\end{cases} \label{eq: chiral BC}
\end{equation}
inducing a zero mode localized at the boundary. \end{enumerate} 

Under (\ref{eq:parity transformation radial}), a solution of the Dirac equation with
angular momentum $m$ obeying boundary condition (\ref{eq:BC Dirac})
will transform into a different solution with angular momentum $-m-1$
obeying (\ref{eq:BC Dirac}) with $h\to-h^{-1}$. Thus, the boundary
condition respects parity if and only if 
\begin{equation}
h_{m}=-h_{-m-1}^{-1}.\label{eq:parity perserving BC}
\end{equation}
 Thus case \ref{list:BC1} above preserves parity while \ref{list:BC2}, \ref{list:BC3}, \ref{list:BC4}
break parity. If (\ref{eq:parity perserving BC}) holds, transformation
(\ref{eq:parity transformation radial}) links between the $m\leftrightarrow-m-1$
eigenspace solutions. The lowest angular momentum subspaces correspond
to orbital angular momentum $m=0,-1$. If (\ref{eq:parity perserving BC})
holds, then the two geometric ladders (\ref{eq: Efimov spectrum-Dirac})
associated with $m=0,-1$ are degenerate. The reason is that, as seen in \eqref{eq: ingoing solution}, under
(\ref{eq:parity transformation radial})
\begin{align}
a_{11} & \to a_{12},\,a_{12}\to-a_{11},\,a_{21}\to a_{22},\,a_{22}\to-a_{21}\nonumber \\
h & \to-h^{-1}\label{eq: parity a}
\end{align}
which render $z_{0}$ in (\ref{eq:z0 explicit}) and consequently
$E_{0}$ invariant. Thus $E_{0,m=\pm1/2}$ are identical in this case.
If (\ref{eq:parity perserving BC}) does not hold, this symmetry is not enforced and the degeneracy
between the ladders is broken. 

The visualization of parity breaking is displayed in Fig. \ref{fig: Dirac efimov ladder} where the density of states $\rho(E)$ of $H_D$ is plotted for the $m=0,-1$ channels and $\beta>\beta_c$. The boundary condition that was used in Fig. \ref{fig: Dirac efimov ladder} is the chiral boundary condition \eqref{eq: chiral BC} which breaks parity. Both curves exhibit an identical set of pronounced peaks condensing near $E=0^-$. These peaks describe quasi-bound states \eqref{eq: Efimov spectrum-Dirac} and, accordingly, are arranged in a set of two geometric ladders. The separation between the ladder is a distinct signal of parity breaking.

\subsection{Experimental realization}

The CSI to DSI transition has recently received further validity and interest due to a detailed experimental observation in graphene \cite{OvdatMaoJiangEtAl2017}. In what follows, we summarize the results of this observation and emphasize its most significant features.

Graphene is a particularly interesting condensed matter system where $H_{D}$ is relevant (for $d=2$). The basic reason for this argument is that low energy excitations in graphene behave as a massless
Dirac fermion field with a linear dispersion $E=\pm v_{F}\left|p\right|$ where the Fermi velocity $v_{F}\approx10^{6}$ m/s appears instead of $c$ \cite{Katsnelson2012d}. These characteristics
have been extensively exploited to make graphene a useful platform
to emulate specific features of quantum field theory, topology and
quantum electrodynamics (QED) \cite{MIRANSKY1980421, ShytovKatsnelsonLevitov2007, ShytovKatsnelsonLevitov2007d,KatsnelsonNovoselovGeim2006,PhysRevLett.102.026807,ZhangTanStormerEtAl2005,WangWongShytovEtAl2013a},
since an effective fine structure constant $\alpha_{G}=e^{2}/\hbar v_{F}$
of order unity is obtained by replacing the velocity of light $c$
by $v_{F}$. 

It has been shown that single-atom vacancies in graphene
can host a local and stable charge \cite{OvdatMaoJiangEtAl2017, mao2016realization,LiuWeinertLi2015}.
This charge can be modified and measured at the vacancy site by
means of scanning tunneling spectroscopy and Landau level spectroscopy
\cite{mao2016realization}. The presence of massless Dirac excitations in the vicinity of the vacancy charge motivates the assumption that these will interact in a way that can be described by a massless Dirac Coulomb system. Particularly, the low energy spectral features of the charged vacancy would be the same as that of a tunable Coulomb source. The experimental results of \cite{OvdatMaoJiangEtAl2017} provide confirmation of this hypothesis as will be detailed below. 

The measurements and data analysis presented below were carried out as follows: positive charges are gradually increased into an initially prepared single atom vacancy in graphene. Using a scanning tunneling microscope (STM) the differential conductance $dI/dV \left( V \right)$ through the STM tip is measured at each charge increment at the vacancy site. The conductance $dI/dV \left( V \right)$ is expected to be proportional to the local density of states of the system \cite{OvdatMaoJiangEtAl2017,akkermans2007mesoscopic}. Thus, quasi-bound states should also appear as pronounced peaks in the $dI/dV$ curves. 

\begin{figure}
\centering
\includegraphics[width=0.8\textwidth]{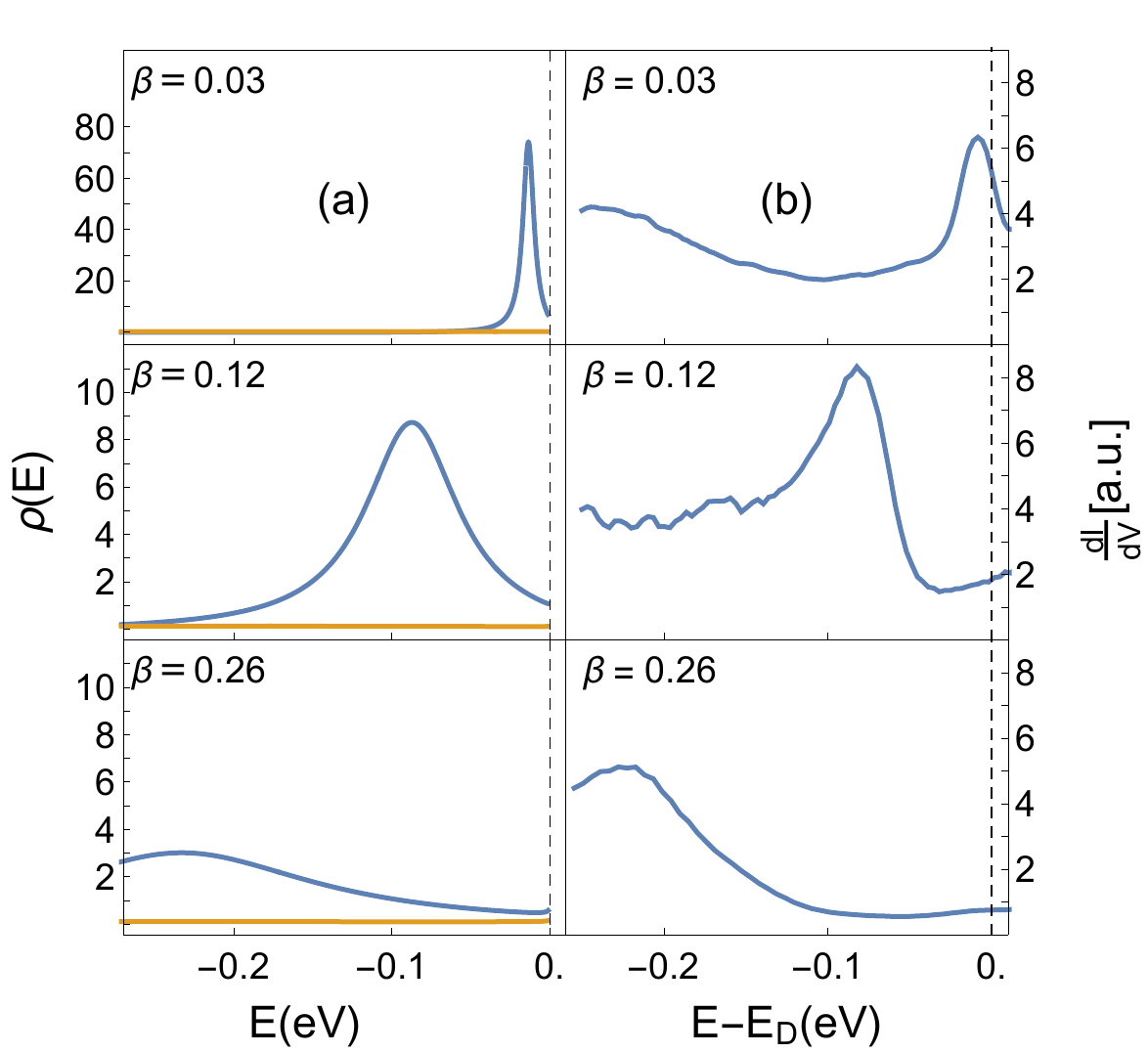}
\caption{\label{fig: under critical Theory Vs Experiment} Experimental and theoretical picture in the undercritical regime. (a) Theoretical behaviour of the density of states $\rho(E)$ of the Dirac Hamiltonian $H_D$ in \eqref{eq: Dirac Hamiltonian} with $d =2$, $c\to v_F = 0.003 c$ and angular momentum channels $m = -1$ (blue) and $m=0$ (yellow). The cutoff and boundary conditions are assigned here with the optimized values $L = 0.195\,\mathrm{nm}$, $h = -0.85(m + 1)$ as explained in the text. The $m = -1$ (blue) branch contains a single peak and the $m = 0$ (yellow) branch shows no peak. While increasing $\beta$, the resonance shifts to lower energy and becomes broader. (b) The conductance $dI/dV$ measured at a single vacancy in graphene using STM as a function of the applied voltage $V$. The determination of the parameter $\beta$ is obtained from matching the position of the peak in the $dI/dV$ curve with the theoretical model where the cutoff $L$ and the boundary condition $h$ are fixed model parameters.}
\end{figure}

\begin{figure}[t]
\centering

\includegraphics[width=1 \textwidth]{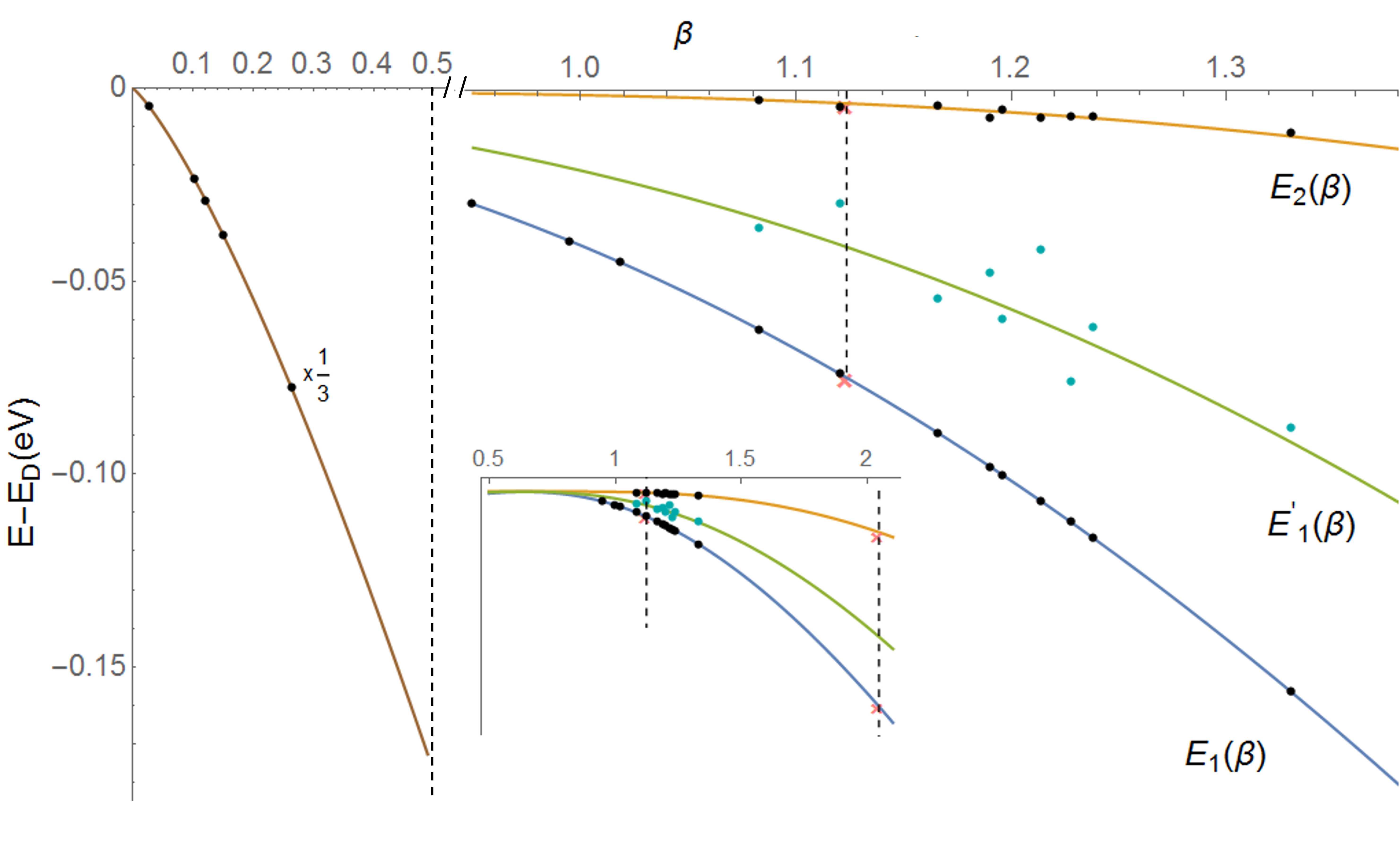}

\caption{\label{fig:CSI to DSI graphene} Comparison of lowest quasi bound state energy curves $E_n (\beta)$ with experimentally measured tunneling conductance peaks. The curves  $E_1(\beta), E'_{1}(\beta), E_2(\beta)$ describe resonances extracted from the density of states of the $m = 0$ ($E_1,\,E_2$) and $m=-1$ ($E'_1$) angular momentum channels. $E_1,\,E_2$ and $E'_1$ are the first quasi bound states appearing for $\beta > 1/2$ in the $m=0,-1$ channels respectively. The brown curve is the position of the single under-critical quasi bound state as a function of $\beta < \beta_c$ scaled by a factor of $1/3$ in the vertical axis. The black and cyan dots correspond to the positions of the tunneling conductance peaks as measured in graphene. The determination of the $\beta$ value associated with these points is obtained from matching the position of the single under-critical peak and first over-critical peak ($E_1$) in the $dI/dV$ curves with the theoretical model where the cutoff $L$ and boundary condition $h$ are fixed parameters. The two pink x's are the values of Efimov energy states as measured in Caesium atoms~\cite{kraemer2006evidence,huang2014observation} and rescaled by an appropriate overall factor. These points corresponds to the (overcritical) fixed Efimov value  $\beta_\mathrm{E} = 1.1236 $. Similarly, additional experimental points obtained in \cite{TungJimenez-GarciaJohansenEtAl2014a, PiresUlmanisHaefnerEtAl2014} are displayed in the inset.}
\end{figure}

For low enough values of the charge, the differential conductance displayed in Fig.~\ref{fig: under critical Theory Vs Experiment}b, shows the existence of a single quasi-bound state resonance whose distance from the Dirac point increases with charge. The behaviour close to the Dirac point, is very similar to the theoretical prediction of the under-critical regime $\beta<\beta_c$ displayed in Fig.~\ref{fig: under critical Theory Vs Experiment}a. The $\beta$ value associated with the data of Fig.~\ref{fig: under critical Theory Vs Experiment}b is obtained from matching the position of the quasi-bound state with the theoretical model where the cutoff $L$ and the boundary condition $h$ are fixed model parameters that will be given later. The theoretical position of the under-critical quasi-bound state as a function of $\beta$ is displayed in Fig. \ref{fig:CSI to DSI graphene} along with the positions of the peak extracted from measurements.  The existence of a quasi-bound state does not contradict CSI of the undercritical phase since the absence of any states occurs only in the low energy limit.

\begin{figure}
\centering
\includegraphics[width=0.8\textwidth]{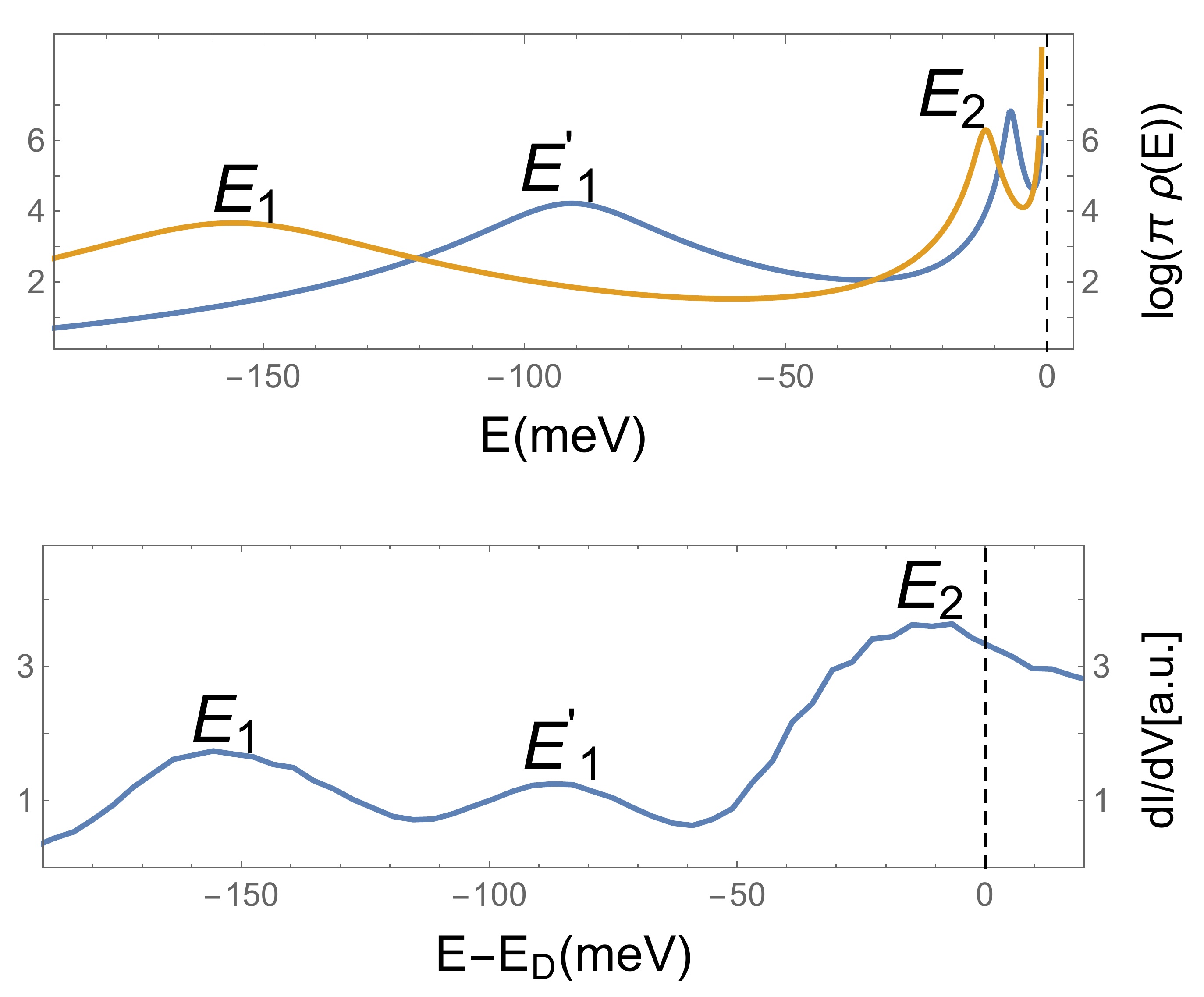}
\caption{ \label{fig: over critical Theory Vs Experiment} Experimental and theoretical picture in the overcritical regime. Upper plot: Theoretical behaviour of the low energy density of states $\rho(E)$ for overcritical $\beta = 1.33$. The Blue (Yellow) line corresponds to $m = -1$ ($m = 0$) orbital angular momentum. The peaks on the vertical scale describe the first quasi-bound states with two (Blue and Yellow) infinite geometric towers of states. Lower plot: Experimental values of the tunnelling conductance measured at the charged vacancy site in graphene.}
\end{figure}

\begin{figure}[t]
\centering

\includegraphics[width=0.8 \textwidth]{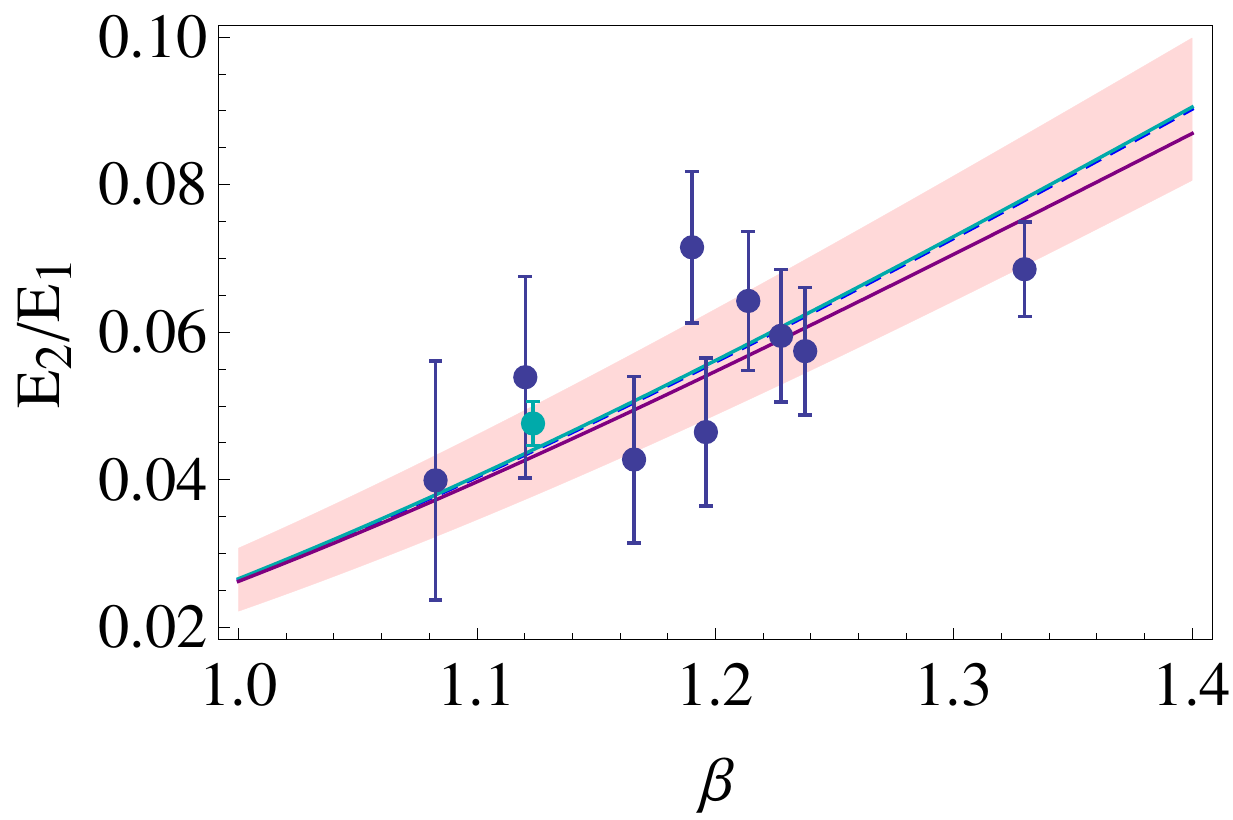}

\caption{\label{fig:theory Vs Exp quantitative} Comparison between the experimentally obtained $E_2/E_1$ ratio and the universal factor $e^{-\pi / \sqrt{\beta^2 - 1/4}}$.
Blue points: the ratio $E_2/E_1$ obtained from the position of the points in Fig.~\ref{fig:CSI to DSI graphene}. Green point: Universal Efimov energy ratio as measured in Caesium atoms~\cite{kraemer2006evidence,huang2014observation}. Blue line (dashed): the corresponding optimized curve, fitted according to the model $e^{-b/\sqrt{\beta^2-1/4}}$ and corresponding to $b = 3.145$ with standard error of $\Delta b = 0.06$ consistent with the predicted value $\pi$. The shaded pink region is the $\pm 2 \Delta b$ confidence interval of the curve. Cyan line: universal low energy factor $e^{-\pi/\sqrt{\beta^2 - 1/4}}$. Purple line: theoretical ratio $E_2/E_1$ obtained from the exact solution of the Dirac equation. As $\beta \rightarrow 0.5 $, $|E_n|$ becomes smaller therefore the green and purple curves coincide for low $\beta$. The error bar on the resonance energies is $\pm 1meV$.}
\end{figure}

At the point where the build up charge exceeds a certain value, three additional resonances emerge out of the Dirac point. These resonances are interpreted as the lowest overcritical $( \beta > 1/2 )$ resonances which we denote  $E_1, E'_{1}, E_2$ respectively. The corresponding theoretical and experimental behaviours displayed in Figs.~\ref{fig: Dirac efimov ladder},~\ref{fig: over critical Theory Vs Experiment}, show a very good qualitative agreement. To achieve a quantitative comparison solely based on the massless Dirac Coulomb Hamiltonian \eqref{eq: 2d H_D}, the theoretical $\beta$ values corresponding to the respective positions of the lowest overcritical experimental resonance $E_1$ (as demonstrated in Fig.~\ref{fig: over critical Theory Vs Experiment}) are deduced for fixed $L$ and the boundary condition $h$ (as before). This allows to determine the lowest branch $E_1 (\beta)$ represented in Fig.~\ref{fig:CSI to DSI graphene}. Then, the experimental points $E'_{1}, E_2$ are now free points to be directly compared to their corresponding theoretical branch as seen in Fig.~\ref{fig:CSI to DSI graphene}. Parameters 
$L$ and $h$, are determined according to the ansatz $h = a(m + 1)$, and correspond to optimal values of $L = 0.195\,\mathrm{nm}$, $a \simeq -0.85$. The comparison of the experimental $E_2/E_1$ ratio with the universal prediction $E_{n+1} / E_n = e^{- \pi / \sqrt{\beta^2 - 1/4}}$ is given in Fig.~\ref{fig:theory Vs Exp quantitative}. A trend-line of the form $e^{-b/\sqrt{\beta^2-1/4}}$ is fitted to the ratios $E_2/E_1$ yielding a statistical value of $b = 3.145$ with standard error of $\Delta b = 0.06$ consistent with the predicted value $\pi$. An error of $\pm 1meV$ is assumed for the position of the energy resonances.

A few further comments are appropriate:
\begin{enumerate}
\item  The points on the $E_2 (\beta)$ curve follow very closely the theoretical prediction $E_{n+1} / E_n = e^{- \pi / \sqrt{\beta^2 - 1/4}}$. This result is relatively insensitive to the choice of $h,L$.
\item  In contrast, the correspondence between the $E'_{1}\left(\beta\right)$ points and the theoretical branch is sensitive to the choice of $h,L$. This reflects the fact that while each geometric ladder is of the form \eqref{eq: Efimov spectrum-Dirac}, the energy scale $E_0$ is different between the $E_1\left(\beta\right)$ and $E'_{1}\left(\beta\right)$ channels thus leading to a shifted relative position. The ansatz taken for $h$ is phenomenological, however, in order to get reasonable correspondence to theory, the explicit dependence on $m$ is needed. More importantly, it is necessary to use a parity breaking boundary condition (see section \ref{subsec: Distinct features associated with spin}) to describe the $E'_{1}(\beta)$ points, otherwise, both angular momentum channels $E_1\left(\beta\right)$, $E'_{1}\left(\beta\right)$ will become degenerate and there would be no theoretical line to describe the $E'_{1}\left(\beta\right)$ points. The existence of the experimental $E'_{1}\left(\beta\right)$ branch is therefore a distinct signal that parity symmetry in the corresponding Dirac description is broken. In graphene, exchanging the triangular sub-lattices is equivalent to a parity transformation. Creating a vacancy breaks the symmetry between the two sub-lattices and is therefore at the origin of broken parity in the Dirac model.
\item The optimal value obtained for the short distance cutoff $L = 0.195 \, \mathrm{nm}$ is fully consistent with the low energy requirement $E_1 L / \hbar v_F \simeq 0.03 \ll 1$ necessary to be in the regime relevant to observe the $\beta$-driven QPT. Furthermore, it is quite close the lattice spacing of graphene ($\approx 0.15 \, \mathrm{nm}$)
\end{enumerate}

One of the most interesting features of observed quasi bound states is their similarity with the Efimov spectrum. As discussed in section \ref{subsec: Realizations of H_S}, Efimov states are a geometric tower of states with a fixed geometric factor which is derived from an effective Schr\"{o}dinger equation with a $V = -\lambda/r^2$ potential (as in \eqref{eq:H_S Schrodinger equation}) and overcritical potential strength $\lambda = s_0 +1/4$, $s_0 \approx 1.00624$. To emphasize the similarities between the Dirac quasi bound spectrum and the Efimov spectrum or, more generally, between the CSI to DSI transition in the Dirac and Schr\"{o}dinger Hamiltonians $H_D$, $H_S$, two additional experimental points (pink x's) are presented in Fig. \ref{fig:CSI to DSI graphene}. These points are the values of Efimov energy states measured in Caesium atoms~\cite{kraemer2006evidence,huang2014observation} and scaled with an appropriate overall factor. The points are placed at the (overcritical) fixed Efimov value $\beta_\mathrm{E} = 1.1236 $ corresponding to the geometric factor of Efimov states. The universality of the transition is thereby emphasized in Fig. \ref{fig:CSI to DSI graphene} in which curves calculated from a massless Dirac Hamiltonian, energy positions of tunneling conductance peaks in graphene and resonances of a gas of Caesium atoms are combined in a meaningful context.

\section{Relation to universality}

In sections \ref{sec:The 1/r^2 -potential}, \ref{sec: Dirac Coulomb system} we obtained the properties of the CSI to DSI transition from a direct analysis of the corresponding eigenstates of each system. In what follows, we describe the same physics, but this time through the language of the renormalization group (RG). As will be detailed next, the description of this phenomenon in a RG picture provides a notable example of a case in which there is universality even in the absence of any fixed points. To understand this point more clearly, we first recall the physical meaning of the RG formalism and the usual context for which universality is understood with relation to RG. 

Universality is a central concept of physics. It refers
to phenomena for which very different systems exhibit identical
behavior when properly coarse-grained to large distance
(or low energy) scales. Important representatives of universality are systems that are close to a critical point, e.g., liquid-gas or magnetic systems. Near the critical point,
these systems exhibit continuous scale invariance (as in \eqref{eq: scaling relation}) where the free energy and correlation
length vary as a power of the temperature (or some other control
parameter). The exponents of these functions are real valued and are
identical for a set of different systems thereby constituting a ``universality
class''.

The contemporary understanding of university in critical phenomena
is provided by the tools of RG and effective
theory. In the framework of the later, low energy physics is described
by a Hamiltonian $H$ with a series of interaction terms $g_{n}\mathcal{O}_{n}$
constrained by symmetries. Intrinsic to this description is an ultraviolet
cutoff $\Lambda$ reflecting the conceptual idea that $H$ is obtained
from some microscopic Hamiltonian $H_{0}$ by integrating out degrees
of freedom with length scale shorter than $1/\Lambda$. The dependence
of $\vec{g}\equiv\left(g_{1},g_{2}\ldots\right)$ on $\Lambda$ defines
the RG space of parameters $\vec{g}\left(\Lambda\right)$ which represent
a large set of Hamiltonians $H\left(\vec{g}\left(\Lambda\right)\right)$.
Within this picture, the scale invariant character of critical phenomena
is attributed to the case where $H_{0}\left(\vec{g}_{0}\right)$ flows
in the infrared limit, $\Lambda\rightarrow0$, to $H\left(\vec{g}_{\ast}\right)$
where $\vec{g}_{\ast}$ is a fixed point. Additionally, universality
classes arise since trajectories starting at distinct positions on RG
space can flow to the same fixed point for $\Lambda\rightarrow0$.
The role of RG fixed points in the description of universality, effective
theory and scale invariance is central and extends throughout broad
sub-fields in physics. 

\subsection{Renomarlization group formalism for the Schr\"{o}dinger $1/r^2$ potential} \label{subsec: RG of 1/r^2}

The RG picture which describes  the low energy physics of the Schr\"{o}dinger $-\lambda/r^2$ potential in the $\lambda>\lambda_{c}$ regime
cannot be associated with a fixed point because of the absence of
CSI. However, even without fixed points, we expect universality to appear in this regime since the geometric series factor
$E_{n+1}/E_{n}$ = $\exp\left(-2\pi/\sqrt{\lambda-\lambda_{c}}\right)$ is independent of the short distance parameters associated with the cutoff $L$ and the boundary condition $g$

To see this explicitly \cite{Kaplan:2009kr,AlbeverioHoegh-KrohnWu1981a,BeaneBedaqueChildressEtAl2001a,MuellerHo2004,KolomeiskyStraley1992c}, consider the radial Schr\"{o}dinger equation for $H_S$ given by
\begin{equation}
-\left(\frac{d^{2}}{dr^{2}}+\frac{d-1}{r}\frac{d}{dr}-\frac{l(l+d-2)}{r^{2}}\right)-\frac{\lambda}{r^{s}} \psi\left(r\right)=E\psi\left(r\right),\hspace{1em}L<r<\infty \label{eq: radial sch equation}
\end{equation}
where $\psi\left(r\right)$ is the radial wavefunction, $l$ the orbital angular momentum, $d$ the space dimension, $L$ a short distance cutoff and $s=2$ but remains implicit for a reason that will be clear shortly. 
A well defined eigenstate of (\ref{eq: radial sch equation}) is obtained by imposing a boundary condition at $r=L$ 
\begin{equation}
L\frac{\psi^{\prime}\left(L\right)}{\psi\left(L\right)}=g, \label{eq:BC RG}
\end{equation}
$g \in \mathbb{R}$, which encodes the short-distance physics. To initiate a RG transformation we transform
\begin{equation}
L\rightarrow L+dL\equiv\epsilon L\,;\hspace{1em}0<\epsilon-1\ll1\label{eq: RG tran}
\end{equation}
and obtain an equivalent effective description with the short distance cut-off $\epsilon L$ and correspondingly, a new boundary condition at $r=\epsilon L$:
\begin{eqnarray}
\epsilon L\frac{\psi^{\prime}\left(\epsilon L\right)}{\psi\left(\epsilon L\right)} & = & g\left(\epsilon L\right).\label{eq:BC trans}
\end{eqnarray}
As a result of \eqref{eq: RG tran}, equation \eqref{eq: radial sch equation} is now defined in the range $\epsilon L\leq r<\infty$ with the same functional form. With the help of the rescaling
$r'\equiv \epsilon^{-1} r,\, E'\equiv\epsilon^{2}E$, equation \eqref{eq: radial sch equation}
is modified to the equivalent form
\begin{equation}
-\left(\frac{d^{2}}{dr'^{2}}+\frac{d-1}{r'}\frac{d}{dr'}-\frac{l(l+d-2)}{r'^{2}}\right)-\frac{\lambda\epsilon^{2-s}}{r'^{s}}\psi\left(r'\right)= E'\psi\left(r'\right)\hspace{1em}L<r'<\infty.    
\end{equation}
Thus, transformation (\ref{eq: RG tran}) is accounted in (\ref{eq: radial sch equation}) by $\lambda \to \lambda \epsilon^{2-s}$ and using (\ref{eq: RG tran}) leads to the infinitesimal form
\begin{equation}
L\frac{d\lambda}{dL}=\left(2-s\right)\lambda.\label{eq: RG eq lambda}
\end{equation}
Similarly, $g\left(\epsilon L\right)$ in (\ref{eq:BC trans}) can be related to $g\left(L\right)$ as follows. The series expansion of $g\left(\epsilon L\right)$ in $\epsilon-1$ is 
\begin{equation}
g\left(\epsilon L\right) = L\frac{\psi^{\prime}\left(L\right)}{\psi\left(L\right)}+\left(\epsilon-1\right)\left(L\frac{\psi^{\prime}\left(L\right)}{\psi\left(L\right)}-L^{2}\left(\frac{\psi^{\prime}\left(L\right)}{\psi\left(L\right)}\right)^{2}+L^{2}\frac{\psi^{\prime\prime}\left(L\right)}{\psi\left(L\right)}\right)+\mathcal{O}\left(\epsilon-1\right)^{2}. \label{eq: g expansion 0}
\end{equation}
Manipulation of \eqref{eq: g expansion 0} by insertion of the radial Schr\"{o}dinger equation \eqref{eq: radial sch equation} and the definition of $g\left(L\right)$ yield 
\begin{equation}
g\left(\epsilon L\right) = g\left(L\right)+\left(\epsilon-1\right)\left(\left(2-d\right)g\left(L\right)-g\left(L\right)^{2}-\lambda L^{2-s}+l(l+d-2)-L^{2} E\right) \label{eq: g expansion}
\end{equation}
where terms of order $\left(\epsilon-1\right)^2$ or higher were eliminated. The equivalent differential form is thus
\begin{equation}
L\frac{dg}{dL}=\left(2-d\right)g-g^{2}-\lambda L^{2-s}+l(l+d-2)-L^{2} E.\label{eq: g RG eq}
\end{equation}
In the low energy regime 
\begin{equation}
L^2\left|E\right|\ll\left|\lambda-l(l+d-2)\right|\label{eq:low E approx}
\end{equation}
equation \eqref{eq: g RG eq} reduces to
\begin{equation}
L\frac{dg}{dL} = \left(2-d\right)g-g^{2}-\lambda\label{eq: g RG eq low E}
\end{equation}
where the orbital angular momentum was taken to be $l=0$ and $s$ set to $s=2$ for brevity. Finally, the combination of \eqref{eq: RG eq lambda}, \eqref{eq: g RG eq low E} constitutes the RG equations
\begin{eqnarray}
\beta\left(\lambda\right) \equiv L\frac{d\lambda}{dL} & = & \left(2-s\right)\lambda \nonumber \\
\beta\left(g\right)  \equiv  L\frac{dg}{dL} & = & -\left(g-g_{+}\right)\left(g-g_{-}\right)\label{eq: RG Schrodinger 1/r^2}
\end{eqnarray}
where 
\begin{equation}
g_{\pm}=\frac{2-d}{2}\pm\sqrt{\lambda_{c}-\lambda}
\end{equation}
and $\lambda_c = \left(d-2\right)^2/4$.

\begin{figure}
\centering
\includegraphics[width=\textwidth]{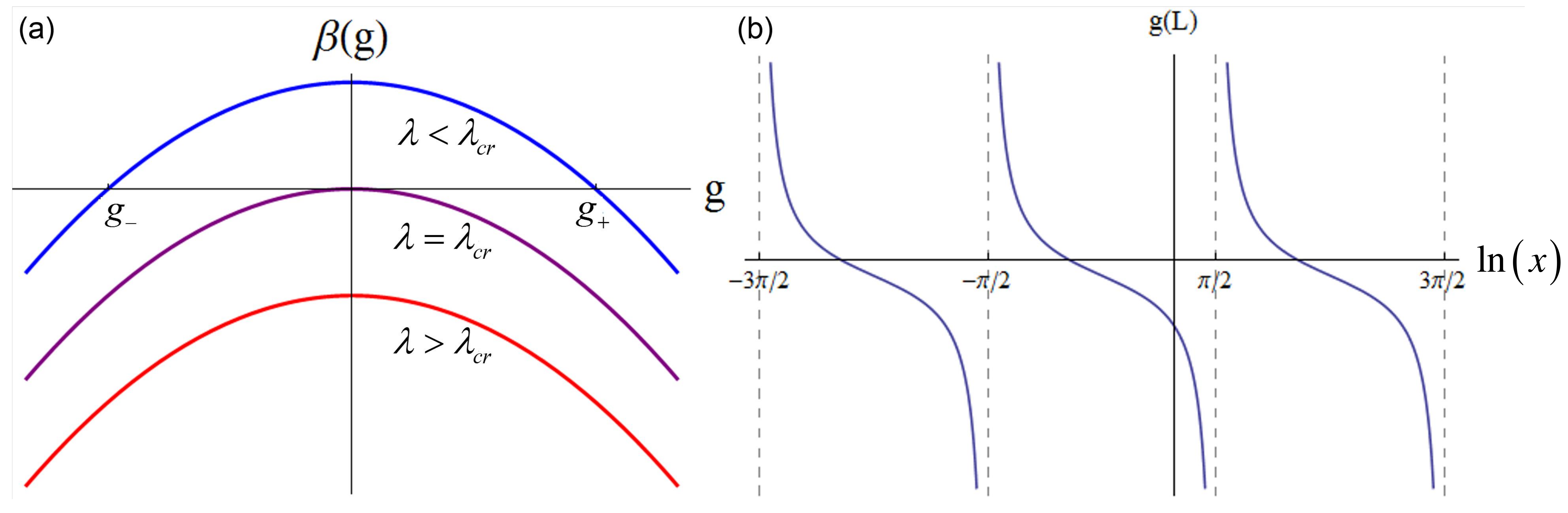}
\caption{\label{fig:RGSchrodinger} Visualization of the renormalization group picture associated with the boundary condition $g(L)$ at the short distance cutoff $r=L$ for the case of the Schr\"{o}dinger $V\left(r\right) = -\lambda/r^2$ potential $H_S$. (a) The $\beta\left(g\right)$ function in the over-critical and
under-critical regimes. For $\lambda < \lambda_{c}$, $\beta\left(g\right)$ has two
roots correspond to two fixed points, $g_-$ unstable and $g_+$
stable. The point $\lambda = \lambda_{c}$ is a transition point where the roots
merge into a single fixed point. For $\lambda > \lambda_{c}$ there are
no real fixed points. (b) The behaviour of the boundary condition $g\left(L\right)$ in the overcritical regime $\lambda > \lambda_{c}$ and $d=3$ as a
function of $\ln \left(x\right)$ with $x \equiv \sqrt{\lambda-\lambda_{c}}\ln\left(L/L_{0}\right) - \phi_{g}$, $\phi_{g} \equiv \arctan\left(\frac{g_{0}-\frac{2-d}{2}}{\sqrt{\lambda-\lambda_{c}}}\right)$. Independent of the initial condition $g_0\left(L_0\right)$, $g\left(L\right)$ is a log-periodic function of $L$ which, as shown in \eqref{eq: DSI scaling}, is a generic feature of DSI.}
\end{figure}

Since $\beta\left(\lambda\right) = 0$ for $s=2$, $\lambda\left(L\right)$ remains unchanged under the RG transformation. In contrary, the function $\beta \left(g\right) $ is not trivial and has two roots $g_{\pm}$ . For $\lambda<\lambda_{c}$,
the two roots correspond to two fixed points, $g_{-}$ unstable and
$g_{+}$ stable. However, as $\lambda$ increases, the two fixed points get closer
and merge for $\lambda=\lambda_{c}$. For $\lambda>\lambda_{c}$,
$g_{\pm}$ become complex valued and the two fixed points vanish as
can be seen in Fig. \ref{fig:RGSchrodinger}a. The solution for $g\left(L\right)$
in this regime is given explicitly by (see Fig. \ref{fig:RGSchrodinger}b)
\begin{equation}
g\left(L\right)=\frac{2-d}{2}-\sqrt{\lambda-\lambda_{c}}\tan\left[\sqrt{\lambda-\lambda_{c}}\ln\left(L/L_{0}\right) - \phi_{g} \right]\label{eq: limit cycle 1/r^2}
\end{equation}
 where $\phi_{g}\equiv\arctan\left(\frac{g_{0}-\frac{2-d}{2}}{\sqrt{\lambda-\lambda_{c}}}\right)$. 
Unlike the case of a fixed point, the flow of $g\left(L\right)$ in \eqref{eq: limit cycle 1/r^2} does not terminate at any specific point but rather oscillate periodically in $\log L$ with period  $L\to e^{\pi/\sqrt{\lambda-\lambda_{c}}}L$ independent of the initial condition $g\left(L_{0}\right) = g_0$. 

The appearance of two fixed points for $\lambda<\lambda_{c}$, which annihilate at $\lambda_{c}$ and give rise to a log-periodic flow for $\lambda>\lambda_{c}$ is the transcription of the CSI to DSI transition in the RG picture. The periodicity $e^{\pi/\sqrt{\lambda-\lambda_{c}}}$, being independent on the initial conditions, $g\left(L_{0}\right) = g_0$, represents a universal content even in the absence of fixed points.

An analogue of the RG equations (\ref{eq: RG Schrodinger 1/r^2}) can be derived for the boundary condition $h\left(L\right)$ in \eqref{eq:BC Dirac} of the massless Dirac Coulomb system described in section \ref{sec: Dirac Coulomb system} \cite{gorsky2014atomic}.

\section{Discussion}

The similarities between the Dirac and Schr\"{o}dinger system $H_{S},H_{D}$
\begin{eqnarray}
H_{S} & = & p^{2}-\lambda/r^{2} \\
H_{D} & = & \gamma^{0}\gamma^{j}p_{j}-\beta/r
\end{eqnarray}
presented in sections \ref{sec:The 1/r^2 -potential}, \ref{sec: Dirac Coulomb system} motivate the study of whether
a similar transition from CSI to DSI is possible for a generic class of systems and,
if so, what are the common ingredients within this class. Below we
briefly survey some other setups which interestingly give rise to a
CSI to DSI transition. The relation between all these cases is summarized
in table \ref{tab:QPTtable summary}.

\begin{table}[t]
\centering{}\caption{\label{tab:QPTtable summary} Comparison between the various cases discussed in the text for which a transition between a continuous scale invariant phase and a discrete scale invariant phase occurs. In the DSI regime each system is characterized by the sudden appearance of a geometric tower of modes with the universal form  $O_{n}=O_{0}\,\exp \left(-b\frac{\pi n}{\sqrt{x-x_{c}}}\right)$. In lines 1--3 of the table, $O_n$ are one body bound states. Lines 4--5 describe many body quantum systems where $O_n$ are fermion masses and 3-body bound states respectively. Line 6 provides a comparison with the Berezinskii-Kosterlitz-Thouless phase transition where the analog quantity for $O_n$ is the free energy $F$ for $T \gtrsim T_c$. In line 4, $\alpha_N$ is a $N$ dependent real number whose exact value can be found in \cite{BrattanPhysRevD.97.061701}. In line 5, $c_{-}$ is a $d$ dependent real positive number defined in section \ref{subsec: Efimov in d dimensions}.}
\medskip{}
\begin{tabular}{lcccc}
\toprule 
System & \quad$O_n$\quad{} & \quad$x$\quad{} & $x_{c}$ &\quad $b$ \quad{} \tabularnewline
\midrule
\addlinespace
$H_{S} = p^2 - \lambda/r^2$ & $E_n$ & $\lambda$ & $\left(d-2\right)^{2}/4$ & $2$\tabularnewline
\addlinespace
$H_{D} = \gamma^{0}\gamma^{j}p_{j}-\beta/r$ & $E_n$ & $\beta^2$ & $\left(d-1\right)^2/4$ & $1$\tabularnewline
\addlinespace
$H_{L} = \left(-\dfrac{d^{2}}{dx^{2}}\right)^{N}-\dfrac{\lambda_{L}}{x^{2N}}$ & $E_n$ & $\lambda_{L}$ & $\left(\dfrac{(2N-1)!!}{2^{N}}\right)^{2}$ & $N \alpha_{N}$\tabularnewline
\addlinespace
QED3 with $N$ massless flavours & $m_n$ & $N^{-1}$ & $\pi^{2}/32$ & $\pi/\sqrt{8}$ \tabularnewline
\addlinespace
Efimov effect in $d$ dimensions & $E_n$ & $d$ & $2.3$ & $1/c_{\pm} \left(d\right)$  \tabularnewline
\addlinespace
BKT & $F$ & $T$ & $T_{c}$ & \text{system dependent} \tabularnewline
\addlinespace
\bottomrule
\end{tabular}
\end{table}

\subsection{Lifshitz scaling symmetry} \label{subsec: Lifshitz}

Since $H_{S}$ and $H_{D}$ share the property that the power law form of
the corresponding potential matches the order of the kinetic term,
it is interesting to examine whether this property is a sufficient
ingredient by considering a generalized class of one dimensional Hamiltonians,
\begin{equation}
H_{L}=\left(-\frac{d^{2}}{dx^{2}}\right)^{N}-\frac{\lambda_{L}}{x^{2N}},\label{Eq: Lifshitz Hamiltonian}
\end{equation}
where $N$ is a natural integer and $\lambda_{L}$ 
a real valued coupling. The Hamiltonian $H_{L}$ describes a quantum system with non-quadratic
anisotropic scaling between space and time for $N>1$. This so called ``Lifshitz
scaling symmetry'' \cite{Alexandre2011}, manifest in (\ref{Eq: Lifshitz Hamiltonian}),
can be seen for example at the finite temperature multicritical points
of certain materials \cite{PhysRevLett.35.1678,PhysRevB.23.4615}
or in strongly correlated electron systems \cite{FradkinHuseMoessnerEtAl2004,VishwanathBalentsSenthil2004,ArdonneFendleyFradkin2004}.
Quartic dispersion relations $E\sim p^{4}$ can also be found in graphene
bilayers \cite{0034-4885-76-5-056503} and heavy fermion metals \cite{2012PhRvL.109q6404R}.
It may also have applications in particle physics \cite{Alexandre2011},
cosmology \cite{Mukohyama2010} and quantum gravity \cite{KachruLiuMulligan2008,Horava2009,Horava2009a}. 

The detailed solution of the corresponding Schr\"{o}dinger equation $H_{L}\psi=E\psi$
\cite{BrattanPhysRevD.97.061701} confirms that a transition from
CSI to DSI occurs at $\lambda_{L,c}=\left(2N-1\right)!!^{2}/2^{2N}$,
$\forall N\geq$1. The CSI phase contains no low energy, $\left|E\right|^{1/2N}L\ll1$ ($L$ is a short distance cutoff),
bound states and the DSI phase is characterized by an infinite set
of bound states forming the geometric series
\begin{equation}
E_{n}=-E_{0}e^{-\frac{N\alpha_{N}\pi n}{\sqrt{\lambda_{L}-\lambda_{L,c}}}},\,\,0<\lambda_{L}-\lambda_{L,c}\ll1\label{eq:Liphsitz spectrum}
\end{equation}
 where $E_0>0$ and $\alpha_{N}$ is an $N$ dependent real number. For $\lambda_{L}-\lambda_{L,c}\to0^{+}$,
the analytic behavior of the spectrum is characteristic of the Berezinskii-Kosterlitz-Thouless (BKT)
scaling in analogy with the $N=1$ case. However, unlike the $N=1$
case, the BKT scaling appears only for $\lambda_{L}-\lambda_{L,c}\to0^{+}$.
Deeper in the overcritical regime, the dependence on $\left(\lambda_{L}-\lambda_{L,c}\right)^{1/2}$
in (\ref{eq:Liphsitz spectrum}) is replaced by a more complicated
function of $\lambda_{L}-\lambda_{L,c}$ \cite{BrattanPhysRevD.97.061701}.
The transition as well as the value of $\lambda_{L,c}$ is independent
of the short distance physics characterized by the boundary conditions
 and cutoff $L$. 

\begin{figure}[t]
\centering

\includegraphics[scale=0.5]{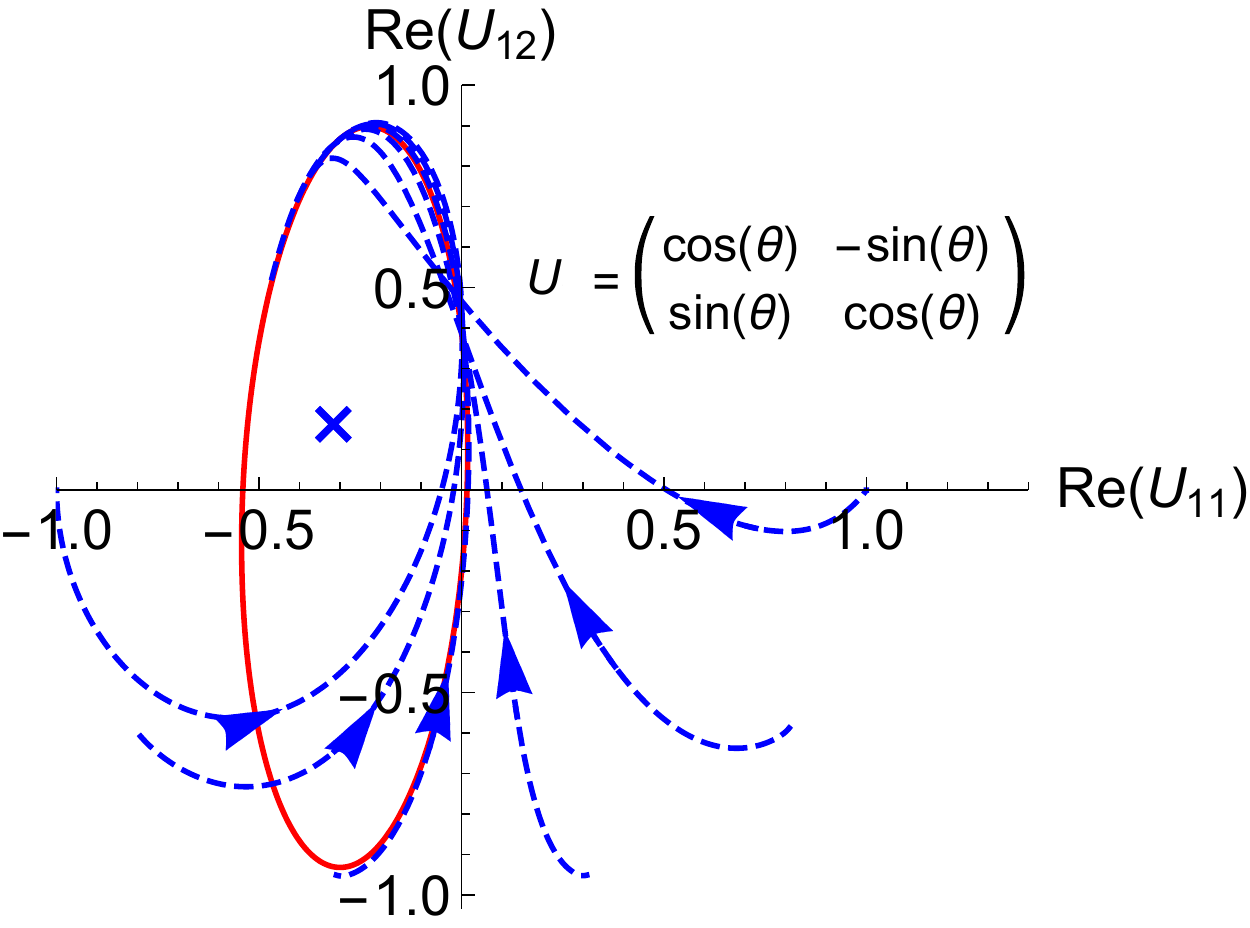}

\caption{\label{fig:isolated limit cycle} A two dimensional projection of the (four dimensional) RG picture of the system $H = d_{x}^{4}-2/x^4$. The four boundary conditions at $x = L$ are parametrized by a unitary $2\times 2$ matrix $U$. The initial conditions for the dashed blue flows are specified by choosing $\theta=-\pi,\ldots,-\pi/10,0$ for the $U$ matrix as displayed. All the trajectories flow towards a limit cycle. There exists a non-unitary fixed point, denoted by the blue cross, which is enclosed by the cycle when projected down onto any two dimensional subspace. }
\end{figure}

Since $H_L$ is a high order differential operator it requires the specification of several $x=L$ boundary condition parameters (unlike the one parameter $g$ in section \ref{subsec: The spectral properties of H_S}) in order to render it as a well defined self-adjoint operator on the interval $L<x<\infty$. The most general choice of boundary conditions is parameterized by a unitary $N\times N$ matrix. Accordingly, the corresponding
$N^{2}$ dimensional RG space is characterized by fixed points
in the under-critical regime $\lambda_{L}<\lambda_{L,c}$. Interestingly, the DSI over-critical regime $\lambda_{L}>\lambda_{L,c}$ is not filled with an infinite number of cyclic flows such as represented in Fig. \ref{fig:RGSchrodinger}b. Instead, there is a 'limit cycle' \cite{strogatz2014nonlinear}, i.e., an isolated closed trajectory at which flows terminate \cite{OvdatJounalOfPhysA} (see Fig. \ref{fig:isolated limit cycle}).

\subsection{QED in $2+1$ dimensions and $N$ fermionic flavors} \label{subsec: QED3}

The study of dynamical fermion mass generation in $2+1$ dimensional quantum electrodynamics (QED) \cite{AppelquistNashWijewardhana1988,HerbutPhysRevD.94.025036} provides an interesting many body instance of the CSI to DSI transition. Consider the $2+1$ dimensional QED Lagrangian 
\begin{equation}
\mathcal{L} = i\bar{\Psi}\gamma^{\mu}\left(\partial_{\mu}-ieA_{\mu}\right)\Psi-\frac{1}{4}F_{\mu\nu}F^{\mu\nu}
\end{equation}
where $\Psi$ is a vector of $N$ identical types of fermion fields with zero mass. In this theory, $e^2$ or alternatively $\alpha \equiv Ne^2/8$, is a dimension-full coupling. Analogous to the short distance cutoff $L$ of sections \ref{sec:The 1/r^2 -potential}, \ref{sec: Dirac Coulomb system}, \ref{subsec: Lifshitz}, $\alpha$ constitutes the only energy scale of the theory. Consequently, the low energy regime $E \ll \alpha$ can be shown to exhibit CSI.

The understand whether or not fermion mass appears as a result of quantum fluctuations it is required to calculate the fermion propagator, specifically, the self-energy $\Sigma \left(p \right)$. Under a particular (non-perturbative) approximation scheme \cite{AppelquistNashWijewardhana1988}, the expression for $\Sigma \left( p \right)$ can be extracted from the solution of the following differential equation
\begin{equation}
    -\Sigma '' \left( p \right) - \frac{2}{p}\Sigma ' \left( p \right) - \frac{\lambda_Q}{p^2+\Sigma \left( p \right)^2} \Sigma \left( p \right) = 0,\hspace{0.5em}0<p<\alpha
    \label{eq: QED3 self energy equation}
\end{equation}
with boundary condition
\begin{equation}
    \alpha \frac{\Sigma'\left(\alpha\right)}{\Sigma\left(\alpha\right)}=-1
    \label{eq: BC Sigma}
\end{equation}
where $\lambda_Q \equiv 8/ \left( \pi^2 N \right)$. Close to a transition point the fermion mass and thereby $\Sigma \left( p \right)$ are non-zero but arbitrarily small such that $\Sigma \left( p \right) \ll p < \alpha$. As a result, \eqref{eq: QED3 self energy equation} can be further approximated by assuming $\Sigma \left( p \right)^2$ in the denominator is a constant which we define as $\Sigma \left( p \right)^2 \to m^2/\lambda_Q$. Expanding to order $m^2$ yields
\begin{equation}
    -\Sigma '' \left( p \right) - \frac{2}{p}\Sigma ' \left( p \right) - \frac{\lambda_Q}{p^2} \Sigma \left( p \right) =  -\frac{m^2}{p^4} \Sigma \left( p \right).
    \label{eq: QED3 self energy equation linearized}
\end{equation}

A closer look on equations \eqref{eq: BC Sigma}, \eqref{eq: QED3 self energy equation linearized} reveals that they are the same as the radial form of the Schr\"{o}dinger equation with a $V= -\lambda/r^2$ potential
\begin{eqnarray}
-\left(\frac{d^{2}}{dr^{2}}+\frac{d-1}{r}\frac{d}{dr}-\frac{l(l+d-2)}{r^{2}}\right)-\frac{\lambda}{r^{2}} \psi\left(r\right) & = & -k^2 \psi\left(r\right),\hspace{0.5em}L<r<\infty \label{eq: Schrodinger 1/r^2 Disscussion} \\
L\frac{\psi^{\prime}\left(L\right)}{\psi\left(L\right)} & = & g \label{eq: BC Schrodinger 1/r^2 Disscussion}
\end{eqnarray}
where $k = \sqrt{-E}$ as described in section \ref{sec:The 1/r^2 -potential} and in equations \eqref{eq: radial sch equation}, \eqref{eq:BC RG}. To see this explicitly, we rewrite \eqref{eq: BC Sigma}, \eqref{eq: QED3 self energy equation linearized} in terms of $r \equiv 1/p$, $\psi\left(r\right) \equiv \Sigma \left(p \right)$, $L \equiv 1/\alpha$ which then yields
\begin{eqnarray}
    -\psi ''(r) -\frac{\lambda_Q }{r^2} \psi (r) & = & - m^2 \psi (r),\hspace{0.5em}L<r<\infty \\
L\frac{\psi^{\prime}\left(L\right)}{\psi\left(L\right)} & = & 1.
    \label{eq: QED3 self energy equation linearized transformed }
\end{eqnarray}
Thus, the appearance of a non-vanishing fermion self energy constitutes a system of the form \eqref{eq: Schrodinger 1/r^2 Disscussion}, \eqref{eq: BC Schrodinger 1/r^2 Disscussion} with $d = 1,\,\lambda = \lambda_Q$ and $g = 1$. The resulting implication is that a transition from a CSI to DSI occurs at $\lambda_{Q,c}=1/4$. For $\lambda_{Q,c}<1/4$ there will be no $\Sigma\left(p\right) \neq 0$ solution for the self-energy in the $m / \alpha \ll 1$ regime. However, once $\lambda_Q$ exceeds $\lambda_{Q,c} = 1/4$ an infinite geometric tower of possible non-trivial self-energy solutions appears with eigenvalues 
\begin{equation}
m_{n+1}/m_n = e^{-\frac{\pi}{\sqrt{\lambda_Q - \lambda_{Q,c}}}}.\label{eq: m_n+1 /m_n}
\end{equation}
The critical point $\lambda_{Q,c}=1/4$ corresponds to a critical fermion number $N_c = 32/\pi^2$ for which the DSI regime is $N<N_c$. In these term, \eqref{eq: m_n+1 /m_n} reduces to 
\begin{equation}
m_{n+1}/m_n = e^{- \frac{\pi}{\sqrt{\frac{1}{N} -\frac{\pi^2}{32}}}\pi/\sqrt{8}}.
\end{equation}

\subsection{Efimov effect in $d$ dimensions} \label{subsec: Efimov in d dimensions}

As described in \ref{subsec: Realizations of H_S}, the Efimov effect \cite{Efimov1970c,Efimov1971b, BraatenHammer2006} is a remarkable phenomenon in which three particles form an infinite geometric ladder of low energy bound states. The effect occurs when at least two of the three pairs interact with a range that is small compared to the scattering length. It can be shown that the Efimov effect is possible only for space dimensions $2.3<d<3.76$ \cite{NIELSEN2001373} which essentially limits the phenomenon to 3 dimensions. Interestingly, in the case where $d$ is allowed to be tuned continuously, two CSI to DSI transitions are initiated at the critical dimensions $d_- = 2.3$, $d_+ = 3.76$ \cite{PhysRevA.98.013633}. In what follows we outline the main features of this result.

\begin{figure}
\centering
\includegraphics[width=\textwidth]{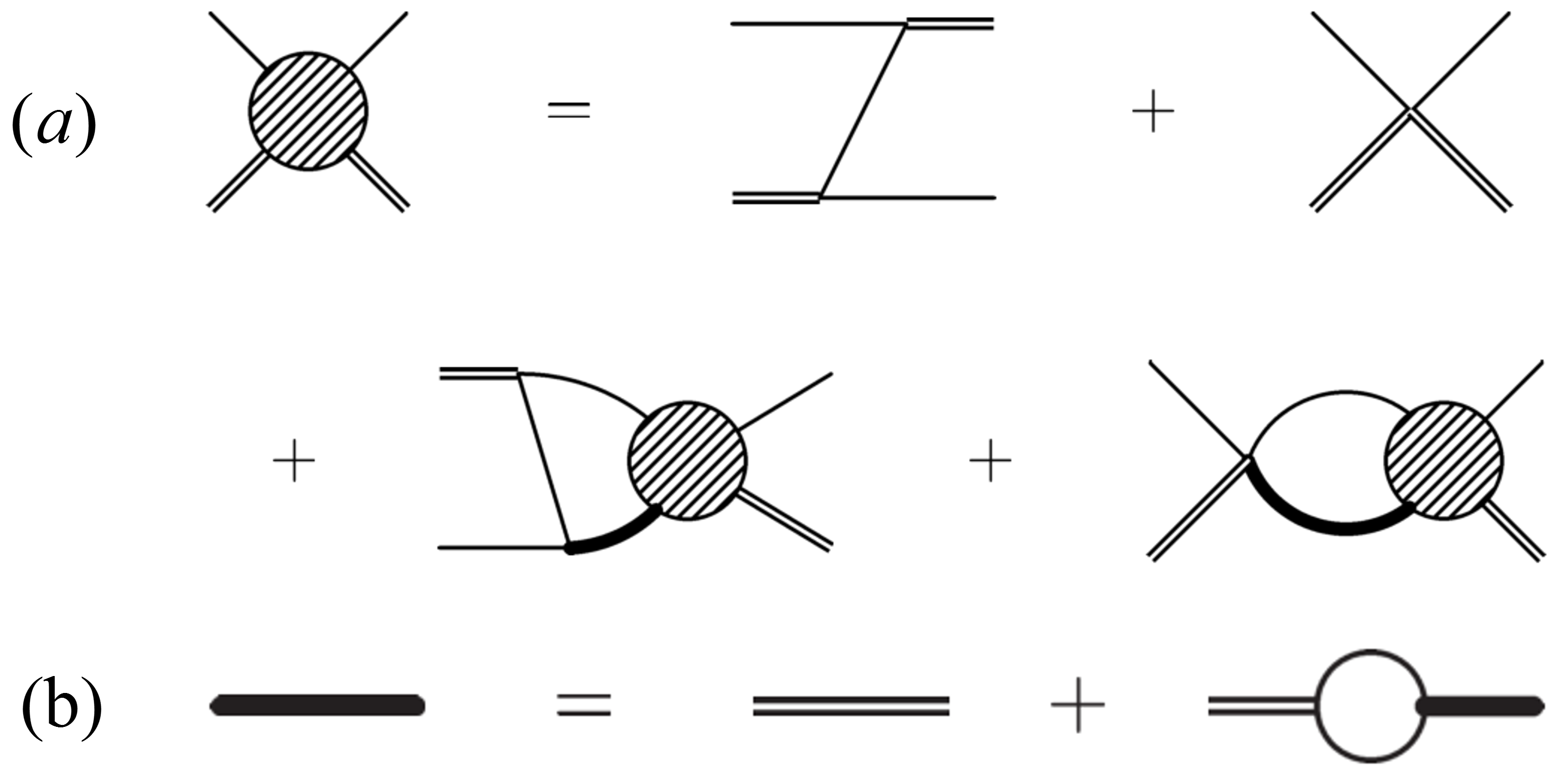}
\caption{\label{fig: Efimov EFT diagrams} Diagrammatic representation of the atom-diatom scattering amplitude and the diatom propagator \cite{BraatenHammer2006}. (a) Diagrammatic self-consistent equation for the atom-diatom scattering amplitude. (b) Diagrammatic self-consistent equation for the diatom field propagator.}
\end{figure}

Low energy 3-body observables of locally interacting identical bosons can be described by an effective field theory with Lagrangian
\begin{equation}
\mathcal{L} = \psi^{\dagger}\left(i\frac{\partial}{\partial t}+\frac{1}{2}\nabla^{2}\right)\psi+\frac{g_{2}}{4}\left(\Delta^{\dagger}\Delta\right)-\frac{g_{2}}{4}\left(\Delta^{\dagger}\psi^{2}+\psi^{\dagger^{2}}\Delta\right)-\frac{g_{3}}{36}\left(\Delta^{\dagger}\Delta\psi^{\dagger}\psi\right)\label{eq:BHvK Lag}
\end{equation}
where $\psi$ is a non-relativistic bosonic atom field, $\Delta$ is a non dynamical 'diatom' field annihilating two atoms at one point 
 and $g_2$, $g_3$ are bare 2-body and 3-body couplings respectively. With the diatom field and these interaction terms, it is possible to reproduce the physics of the Efimov effect \cite{BedaqueHammerVanKloPhysRevLett.82.463}. The main ingredient of this procedure is the diagramatic calculation of the atom-diatom scattering amplitude as shown in Fig. \ref{fig: Efimov EFT diagrams}. The self-consistent equations described in Fig. \ref{fig: Efimov EFT diagrams} leads to the following approximate relation for the s-wave atom-diatom amplitude $A_s$
\begin{equation}
   A_s \left( p \right) =  -\left(\frac{4}{3}\right)^{\frac{d-2}{2}} \frac{4\sin\left(\frac{d}{2} \pi \right)}{\pi} \int_0^\infty\, dq \frac{q}{p^2+q^2} \,  {}_2 F_1\left(\begin{matrix}\frac{1}{2}& &1 \\ &\frac{d}{2}\end{matrix};\frac{p^2 q^2}{p^2+q^2} \right)    A_s\left( q \right)
   \label{eq: amplitude CSi relation}
\end{equation}
Since there are no dimension-full parameters in \eqref{eq: amplitude CSi relation} we are, once again, faced with a CSI equation, in analogy with the characteristics of equations \eqref{eq: 1/r^2 Schrodinger}, \eqref{eq: Dirac equation}, \eqref{Eq: Lifshitz Hamiltonian} and \eqref{eq: QED3 self energy equation linearized}. By inserting the ansatz $A_s \left( p \right) = p^{s-1}$, two possible solutions for \eqref{eq: amplitude CSi relation} are obtained
\begin{equation}
    A_s \approx a_1 p^{\sqrt{s^2}-1} + a_2 p^{-\sqrt{s^2}-1} \label{eq: A_s solution}
\end{equation}
where $s^2 \left( d \right)$ is a solution of the $s\to-s$ invariant equation
\begin{equation}
2 \sin \left( \frac{d\pi}{2}\right) {}_2 F_1\left(\begin{matrix}\frac{d-1+s}{2}& &\frac{d-1-s}{2} \\ &\frac{d}{2}\end{matrix};\frac{1}{4} \right)+ \cos \left( \frac{s}{2} \pi \right) = 0. \label{eq: s transendal equation}
\end{equation}
The numerical solution $s^2 \left(d \right)$ of \eqref{eq: s transendal equation} shows that near $d=d_{\pm}$, $s^2 \left( d_\pm \right) = 0$, $\partial_d s^{2} \left(d_-\right) < 0$, $\partial_d s^{2} \left(d_+\right) > 0$ and it is analytic. Consequently, near the critical dimensions $d_\pm$
\begin{equation}
    s^2\left( d\right) = \pm c_\pm^2 \left(d-d_\pm\right)+\mathcal{O}\left(d-d_{\pm}\right)
    \label{eq: near critical dimension s^2}
\end{equation}
with $c_\pm > 0$. The insertion of \eqref{eq: near critical dimension s^2} into \eqref{eq: A_s solution} imply a CSI to DSI transition from real to complex valued power law behaviour of $A_s$. The DSI regime $d_-<d<d_+$ is consistent with the strip within which the Efimov states appear. Consequently, close to the critical points $d=d_{\pm}$, $A_s \left(p\right)$ in \eqref{eq: A_s solution} obeys the following DSI scaling relation (as in \eqref{eq: scaling relation})
\begin{equation}
    A_s \left(e^{\frac{\pi n}{c_{\pm}\sqrt{|d-d_{\pm}|}}}p\right) = e^{-\frac{\pi n}{c_{\pm}\sqrt{|d-d_{\pm}|}}}  A_s \left(p\right).
\end{equation}
The corresponding RG equation for the couplings is
\begin{equation}
    \Lambda \frac{d}{d\Lambda} G = \frac{1 - s^2 \left( d \right)}{2} \left(G - G_- \right)\left(G - G_+ \right)
\end{equation}
where $G\left( \Lambda \right) \equiv  \Lambda ^2 g_3 \left(\Lambda\right)/9 g_2 \left(\Lambda\right)^2$, $\Lambda$ is a UV cutoff and 
\begin{equation}
G_\pm \equiv - \left(1\pm \sqrt{s^2\left( d \right)}\right)/\left(1\mp \sqrt{s^2\left( d \right)}\right).    
\end{equation}
In accordance with the RG picture detailed in section \ref{subsec: RG of 1/r^2}, the insertion of \eqref{eq: near critical dimension s^2} shows that the $\beta$-function of $G$ contains two fixed points outside the strip $d_- < d < d_+ $ which annihilate at $d=d_\pm$.

\section{Summary}

The breaking of continuous scale invariance (CSI) into discrete scale invariance (DSI) is a rich phenomenon with roots in multiple fields in physics. Theoretically, this transition plays an important role in various fundamental quantum systems such as the inverse-squared potential (section \ref{sec:The 1/r^2 -potential}), the massless hydrogen atom (section \ref{sec: Dirac Coulomb system}), $2+1$ dimensional quantum electrodynamics (section \ref{subsec: QED3}) and the Efimov effect (sections \ref{subsec: Realizations of H_S} and \ref{subsec: Efimov in d dimensions}). This CSI to DSI transition constitutes a quantum phase transition which appears for single body and strongly coupled many body systems and extends through non-relativistic, relativistic and Lifshitz dispersion relations. In a RG picture the transition describes universal low energy physics without fixed points and constitutes a physical realization of a limit cycle. Remarkably, the features of this transition have been measured recently in various systems such as cold atoms, graphene and Fermi gases \cite{Deng2016}. In the DSI phase, the dependence of the geometric ladder of states on the control parameter (see Table \ref{tab:QPTtable summary}) is in the class of Berezinskii-Kosterlitz-Thouless transitions. This provides an interesting, yet to be studied, bridge between DSI and two dimensional systems associated with BKT physics. 

The characteristics described above provide the motivation to further study the ingredients associated with CSI to DSI transitions and we expect that these transitions will have an increasingly important role across the physics community in the future.

\section*{Acknowledgements}
This work was supported by the Israel Science Foundation Grant No.~924/09 and by the Pazy Foundation.
\bibliography{refs}

\begin{thebibliography}{71}%
\makeatletter
\providecommand \@ifxundefined [1]{%
 \@ifx{#1\undefined}
}%
\providecommand \@ifnum [1]{%
 \ifnum #1\expandafter \@firstoftwo
 \else \expandafter \@secondoftwo
 \fi
}%
\providecommand \@ifx [1]{%
 \ifx #1\expandafter \@firstoftwo
 \else \expandafter \@secondoftwo
 \fi
}%
\providecommand \natexlab [1]{#1}%
\providecommand \enquote  [1]{``#1''}%
\providecommand \bibnamefont  [1]{#1}%
\providecommand \bibfnamefont [1]{#1}%
\providecommand \citenamefont [1]{#1}%
\providecommand \href@noop [0]{\@secondoftwo}%
\providecommand \href [0]{\begingroup \@sanitize@url \@href}%
\providecommand \@href[1]{\@@startlink{#1}\@@href}%
\providecommand \@@href[1]{\endgroup#1\@@endlink}%
\providecommand \@sanitize@url [0]{\catcode `\\12\catcode `\$12\catcode
  `\&12\catcode `\#12\catcode `\^12\catcode `\_12\catcode `\%12\relax}%
\providecommand \@@startlink[1]{}%
\providecommand \@@endlink[0]{}%
\providecommand \url  [0]{\begingroup\@sanitize@url \@url }%
\providecommand \@url [1]{\endgroup\@href {#1}{\urlprefix }}%
\providecommand \urlprefix  [0]{URL }%
\providecommand \Eprint [0]{\href }%
\providecommand \doibase [0]{http://dx.doi.org/}%
\providecommand \selectlanguage [0]{\@gobble}%
\providecommand \bibinfo  [0]{\@secondoftwo}%
\providecommand \bibfield  [0]{\@secondoftwo}%
\providecommand \translation [1]{[#1]}%
\providecommand \BibitemOpen [0]{}%
\providecommand \bibitemStop [0]{}%
\providecommand \bibitemNoStop [0]{.\EOS\space}%
\providecommand \EOS [0]{\spacefactor3000\relax}%
\providecommand \BibitemShut  [1]{\csname bibitem#1\endcsname}%
\let\auto@bib@innerbib\@empty
\bibitem [{\citenamefont {Case}(1950)}]{Case1950b}%
  \BibitemOpen
  \bibfield  {author} {\bibinfo {author} {\bibfnamefont {K.~M.}\ \bibnamefont
  {Case}},\ }\href {\doibase 10.1103/PhysRev.80.797} {\bibfield  {journal}
  {\bibinfo  {journal} {Phys. Rev.}\ }\textbf {\bibinfo {volume} {80}},\
  \bibinfo {pages} {797} (\bibinfo {year} {1950})}\BibitemShut {NoStop}%
\bibitem [{\citenamefont {Landau}(1991)}]{Landau1991b}%
  \BibitemOpen
  \bibfield  {author} {\bibinfo {author} {\bibfnamefont {L.~D.}\ \bibnamefont
  {Landau}},\ }\href@noop {} {\emph {\bibinfo {title} {Quantum mechanics :
  non-relativistic theory}}}\ (\bibinfo  {publisher} {Butterworth-Heinemann},\
  \bibinfo {address} {Oxford Boston},\ \bibinfo {year} {1991})\BibitemShut
  {NoStop}%
\bibitem [{\citenamefont {Efimov}(1970)}]{Efimov1970c}%
  \BibitemOpen
  \bibfield  {author} {\bibinfo {author} {\bibfnamefont {V.}~\bibnamefont
  {Efimov}},\ }\href {\doibase 10.1016/0370-2693(70)90349-7} {\bibfield
  {journal} {\bibinfo  {journal} {Physics Letters B}\ }\textbf {\bibinfo
  {volume} {33}},\ \bibinfo {pages} {563} (\bibinfo {year} {1970})}\BibitemShut
  {NoStop}%
\bibitem [{\citenamefont {Efimov}(1971)}]{Efimov1971b}%
  \BibitemOpen
  \bibfield  {author} {\bibinfo {author} {\bibfnamefont {V.}~\bibnamefont
  {Efimov}},\ }\href
  {https://www.uibk.ac.at/exphys/ultracold/projects/levt/FourBodies/SovJNucPhys12.589.efimov.pdf}
  {\bibfield  {journal} {\bibinfo  {journal} {Sov. J. Nucl. Phys}\ }\textbf
  {\bibinfo {volume} {12}},\ \bibinfo {pages} {589} (\bibinfo {year}
  {1971})}\BibitemShut {NoStop}%
\bibitem [{\citenamefont {L\'evy-Leblond}(1967)}]{Levy-Leblond1967}%
  \BibitemOpen
  \bibfield  {author} {\bibinfo {author} {\bibfnamefont {J.-M.}\ \bibnamefont
  {L\'evy-Leblond}},\ }\href {\doibase 10.1103/PhysRev.153.1} {\bibfield
  {journal} {\bibinfo  {journal} {Phys. Rev.}\ }\textbf {\bibinfo {volume}
  {153}},\ \bibinfo {pages} {1} (\bibinfo {year} {1967})}\BibitemShut {NoStop}%
\bibitem [{\citenamefont {Camblong}\ \emph {et~al.}(2001)\citenamefont
  {Camblong}, \citenamefont {Epele}, \citenamefont {Fanchiotti},\ and\
  \citenamefont {Garcia~Canal}}]{Camblong:2001zt}%
  \BibitemOpen
  \bibfield  {author} {\bibinfo {author} {\bibfnamefont {H.~E.}\ \bibnamefont
  {Camblong}}, \bibinfo {author} {\bibfnamefont {L.~N.}\ \bibnamefont {Epele}},
  \bibinfo {author} {\bibfnamefont {H.}~\bibnamefont {Fanchiotti}}, \ and\
  \bibinfo {author} {\bibfnamefont {C.~A.}\ \bibnamefont {Garcia~Canal}},\
  }\href {\doibase 10.1103/PhysRevLett.87.220402} {\bibfield  {journal}
  {\bibinfo  {journal} {Phys.Rev.Lett.}\ }\textbf {\bibinfo {volume} {87}},\
  \bibinfo {pages} {220402} (\bibinfo {year} {2001})},\ \Eprint
  {http://arxiv.org/abs/hep-th/0106144} {arXiv:hep-th/0106144 [hep-th]}
  \BibitemShut {NoStop}%
\bibitem [{\citenamefont {Kaplan}\ \emph {et~al.}(2009)\citenamefont {Kaplan},
  \citenamefont {Lee}, \citenamefont {Son},\ and\ \citenamefont
  {Stephanov}}]{Kaplan:2009kr}%
  \BibitemOpen
  \bibfield  {author} {\bibinfo {author} {\bibfnamefont {D.~B.}\ \bibnamefont
  {Kaplan}}, \bibinfo {author} {\bibfnamefont {J.-W.}\ \bibnamefont {Lee}},
  \bibinfo {author} {\bibfnamefont {D.~T.}\ \bibnamefont {Son}}, \ and\
  \bibinfo {author} {\bibfnamefont {M.~A.}\ \bibnamefont {Stephanov}},\ }\href
  {\doibase 10.1103/PhysRevD.80.125005} {\bibfield  {journal} {\bibinfo
  {journal} {Phys. Rev. D}\ }\textbf {\bibinfo {volume} {80}},\ \bibinfo
  {pages} {125005} (\bibinfo {year} {2009})}\BibitemShut {NoStop}%
\bibitem [{\citenamefont {Nisoli}\ and\ \citenamefont
  {Bishop}(2014)}]{NisoliBishop2014}%
  \BibitemOpen
  \bibfield  {author} {\bibinfo {author} {\bibfnamefont {C.}~\bibnamefont
  {Nisoli}}\ and\ \bibinfo {author} {\bibfnamefont {A.~R.}\ \bibnamefont
  {Bishop}},\ }\href {\doibase 10.1103/PhysRevLett.112.070401} {\bibfield
  {journal} {\bibinfo  {journal} {Phys. Rev. Lett.}\ }\textbf {\bibinfo
  {volume} {112}},\ \bibinfo {pages} {070401} (\bibinfo {year}
  {2014})}\BibitemShut {NoStop}%
\bibitem [{\citenamefont {De~Martino}\ \emph {et~al.}(2014)\citenamefont
  {De~Martino}, \citenamefont {Kl\"opfer}, \citenamefont {Matrasulov},\ and\
  \citenamefont {Egger}}]{DeMartinoKloepferMatrasulovEtAl2014}%
  \BibitemOpen
  \bibfield  {author} {\bibinfo {author} {\bibfnamefont {A.}~\bibnamefont
  {De~Martino}}, \bibinfo {author} {\bibfnamefont {D.}~\bibnamefont
  {Kl\"opfer}}, \bibinfo {author} {\bibfnamefont {D.}~\bibnamefont
  {Matrasulov}}, \ and\ \bibinfo {author} {\bibfnamefont {R.}~\bibnamefont
  {Egger}},\ }\href {\doibase 10.1103/PhysRevLett.112.186603} {\bibfield
  {journal} {\bibinfo  {journal} {Phys. Rev. Lett.}\ }\textbf {\bibinfo
  {volume} {112}},\ \bibinfo {pages} {186603} (\bibinfo {year}
  {2014})}\BibitemShut {NoStop}%
\bibitem [{\citenamefont {Jackiw}(1995)}]{Jackiw1995b}%
  \BibitemOpen
  \bibfield  {author} {\bibinfo {author} {\bibfnamefont {R.~W.}\ \bibnamefont
  {Jackiw}},\ }\href@noop {} {\emph {\bibinfo {title} {Diverse topics in
  theoretical and mathematical physics}}}\ (\bibinfo  {publisher} {World
  Scientific},\ \bibinfo {year} {1995})\BibitemShut {NoStop}%
\bibitem [{\citenamefont {Meetz}(1964)}]{Meetz1964d}%
  \BibitemOpen
  \bibfield  {author} {\bibinfo {author} {\bibfnamefont {K.}~\bibnamefont
  {Meetz}},\ }\href {\doibase 10.1007/BF02750010} {\bibfield  {journal}
  {\bibinfo  {journal} {Il Nuovo Cimento (1955-1965)}\ }\textbf {\bibinfo
  {volume} {34}},\ \bibinfo {pages} {690} (\bibinfo {year} {1964})}\BibitemShut
  {NoStop}%
\bibitem [{\citenamefont {Gitman}\ \emph {et~al.}(2012)\citenamefont {Gitman},
  \citenamefont {Tyutin},\ and\ \citenamefont
  {Voronov}}]{GitmanTyutinVoronov2012c}%
  \BibitemOpen
  \bibfield  {author} {\bibinfo {author} {\bibfnamefont {D.~M.}\ \bibnamefont
  {Gitman}}, \bibinfo {author} {\bibfnamefont {I.}~\bibnamefont {Tyutin}}, \
  and\ \bibinfo {author} {\bibfnamefont {B.~L.}\ \bibnamefont {Voronov}},\
  }\href@noop {} {\emph {\bibinfo {title} {Self-adjoint Extensions in Quantum
  Mechanics: General Theory and Applications to Schr{\"o}dinger and Dirac
  Equations with Singular Potentials}}},\ Vol.~\bibinfo {volume} {62}\
  (\bibinfo  {publisher} {Springer},\ \bibinfo {year} {2012})\BibitemShut
  {NoStop}%
\bibitem [{\citenamefont {Albeverio}\ \emph {et~al.}(1981)\citenamefont
  {Albeverio}, \citenamefont {H{\o}egh-Krohn},\ and\ \citenamefont
  {Wu}}]{AlbeverioHoegh-KrohnWu1981a}%
  \BibitemOpen
  \bibfield  {author} {\bibinfo {author} {\bibfnamefont {S.}~\bibnamefont
  {Albeverio}}, \bibinfo {author} {\bibfnamefont {R.}~\bibnamefont
  {H{\o}egh-Krohn}}, \ and\ \bibinfo {author} {\bibfnamefont {T.~T.}\
  \bibnamefont {Wu}},\ }\href {\doibase 10.1016/0375-9601(81)90507-7}
  {\bibfield  {journal} {\bibinfo  {journal} {Phys. Lett. A}\ }\textbf
  {\bibinfo {volume} {83}},\ \bibinfo {pages} {105} (\bibinfo {year}
  {1981})}\BibitemShut {NoStop}%
\bibitem [{\citenamefont {Beane}\ \emph {et~al.}(2001)\citenamefont {Beane},
  \citenamefont {Bedaque}, \citenamefont {Childress}, \citenamefont
  {Kryjevski}, \citenamefont {McGuire},\ and\ \citenamefont {van
  Kolck}}]{BeaneBedaqueChildressEtAl2001a}%
  \BibitemOpen
  \bibfield  {author} {\bibinfo {author} {\bibfnamefont {S.~R.}\ \bibnamefont
  {Beane}}, \bibinfo {author} {\bibfnamefont {P.~F.}\ \bibnamefont {Bedaque}},
  \bibinfo {author} {\bibfnamefont {L.}~\bibnamefont {Childress}}, \bibinfo
  {author} {\bibfnamefont {A.}~\bibnamefont {Kryjevski}}, \bibinfo {author}
  {\bibfnamefont {J.}~\bibnamefont {McGuire}}, \ and\ \bibinfo {author}
  {\bibfnamefont {U.}~\bibnamefont {van Kolck}},\ }\href {\doibase
  10.1103/PhysRevA.64.042103} {\bibfield  {journal} {\bibinfo  {journal} {Phys.
  Rev. A}\ }\textbf {\bibinfo {volume} {64}},\ \bibinfo {pages} {042103}
  (\bibinfo {year} {2001})}\BibitemShut {NoStop}%
\bibitem [{\citenamefont {Mueller}\ and\ \citenamefont
  {Ho}(2004)}]{MuellerHo2004}%
  \BibitemOpen
  \bibfield  {author} {\bibinfo {author} {\bibfnamefont {E.~J.}\ \bibnamefont
  {Mueller}}\ and\ \bibinfo {author} {\bibfnamefont {T.-L.}\ \bibnamefont
  {Ho}},\ }\href {https://arxiv.org/pdf/cond-mat/0403283.pdf} {\bibfield
  {journal} {\bibinfo  {journal} {arXiv preprint cond-mat/0403283}\ } (\bibinfo
  {year} {2004})}\BibitemShut {NoStop}%
\bibitem [{\citenamefont {Braaten}\ and\ \citenamefont
  {Phillips}(2004)}]{BraatenPhillips2004b}%
  \BibitemOpen
  \bibfield  {author} {\bibinfo {author} {\bibfnamefont {E.}~\bibnamefont
  {Braaten}}\ and\ \bibinfo {author} {\bibfnamefont {D.}~\bibnamefont
  {Phillips}},\ }\href {\doibase 10.1103/PhysRevA.70.052111} {\bibfield
  {journal} {\bibinfo  {journal} {Phys. Rev. A}\ }\textbf {\bibinfo {volume}
  {70}},\ \bibinfo {pages} {052111} (\bibinfo {year} {2004})}\BibitemShut
  {NoStop}%
\bibitem [{\citenamefont {Hammer}\ and\ \citenamefont
  {Swingle}(2006)}]{HammerSwingle2006c}%
  \BibitemOpen
  \bibfield  {author} {\bibinfo {author} {\bibfnamefont {H.-W.}\ \bibnamefont
  {Hammer}}\ and\ \bibinfo {author} {\bibfnamefont {B.~G.}\ \bibnamefont
  {Swingle}},\ }\href {\doibase 10.1016/j.aop.2005.04.017} {\bibfield
  {journal} {\bibinfo  {journal} {Annals of Physics}\ }\textbf {\bibinfo
  {volume} {321}},\ \bibinfo {pages} {306} (\bibinfo {year}
  {2006})}\BibitemShut {NoStop}%
\bibitem [{\citenamefont {Moroz}\ and\ \citenamefont
  {Schmidt}(2010)}]{MorozSchmidt2009}%
  \BibitemOpen
  \bibfield  {author} {\bibinfo {author} {\bibfnamefont {S.}~\bibnamefont
  {Moroz}}\ and\ \bibinfo {author} {\bibfnamefont {R.}~\bibnamefont
  {Schmidt}},\ }\href {\doibase 10.1016/j.aop.2009.10.002} {\bibfield
  {journal} {\bibinfo  {journal} {Annals of Physics}\ }\textbf {\bibinfo
  {volume} {325}},\ \bibinfo {pages} {491} (\bibinfo {year}
  {2010})}\BibitemShut {NoStop}%
\bibitem [{\citenamefont {Kolomeisky}\ and\ \citenamefont
  {Straley}(1992)}]{KolomeiskyStraley1992c}%
  \BibitemOpen
  \bibfield  {author} {\bibinfo {author} {\bibfnamefont {E.~B.}\ \bibnamefont
  {Kolomeisky}}\ and\ \bibinfo {author} {\bibfnamefont {J.~P.}\ \bibnamefont
  {Straley}},\ }\href {\doibase 10.1103/PhysRevB.46.12664} {\bibfield
  {journal} {\bibinfo  {journal} {Phys. Rev. B}\ }\textbf {\bibinfo {volume}
  {46}},\ \bibinfo {pages} {12664} (\bibinfo {year} {1992})}\BibitemShut
  {NoStop}%
\bibitem [{\citenamefont {Jensen}(2011)}]{Jensen2011}%
  \BibitemOpen
  \bibfield  {author} {\bibinfo {author} {\bibfnamefont {K.}~\bibnamefont
  {Jensen}},\ }\href {\doibase 10.1103/PhysRevLett.107.231601} {\bibfield
  {journal} {\bibinfo  {journal} {Phys. Rev. Lett.}\ }\textbf {\bibinfo
  {volume} {107}},\ \bibinfo {pages} {231601} (\bibinfo {year} {2011})},\
  \Eprint {http://arxiv.org/abs/1108.0421} {arXiv:1108.0421 [hep-th]}
  \BibitemShut {NoStop}%
\bibitem [{\citenamefont {Jensen}\ \emph {et~al.}(2010)\citenamefont {Jensen},
  \citenamefont {Karch}, \citenamefont {Son},\ and\ \citenamefont
  {Thompson}}]{JensenKarchSonEtAl2010}%
  \BibitemOpen
  \bibfield  {author} {\bibinfo {author} {\bibfnamefont {K.}~\bibnamefont
  {Jensen}}, \bibinfo {author} {\bibfnamefont {A.}~\bibnamefont {Karch}},
  \bibinfo {author} {\bibfnamefont {D.~T.}\ \bibnamefont {Son}}, \ and\
  \bibinfo {author} {\bibfnamefont {E.~G.}\ \bibnamefont {Thompson}},\ }\href
  {\doibase 10.1103/PhysRevLett.105.041601} {\bibfield  {journal} {\bibinfo
  {journal} {Phys. Rev. Lett.}\ }\textbf {\bibinfo {volume} {105}},\ \bibinfo
  {pages} {041601} (\bibinfo {year} {2010})},\ \Eprint
  {http://arxiv.org/abs/1002.3159} {arXiv:1002.3159 [hep-th]} \BibitemShut
  {NoStop}%
\bibitem [{\citenamefont {Braaten}\ and\ \citenamefont
  {Hammer}(2006)}]{BraatenHammer2006}%
  \BibitemOpen
  \bibfield  {author} {\bibinfo {author} {\bibfnamefont {E.}~\bibnamefont
  {Braaten}}\ and\ \bibinfo {author} {\bibfnamefont {H.-W.}\ \bibnamefont
  {Hammer}},\ }\href {\doibase https://doi.org/10.1016/j.physrep.2006.03.001}
  {\bibfield  {journal} {\bibinfo  {journal} {Physics Reports}\ }\textbf
  {\bibinfo {volume} {428}},\ \bibinfo {pages} {259 } (\bibinfo {year}
  {2006})}\BibitemShut {NoStop}%
\bibitem [{\citenamefont {Kraemer}\ \emph {et~al.}(2006)\citenamefont
  {Kraemer}, \citenamefont {Mark}, \citenamefont {Waldburger}, \citenamefont
  {Danzl}, \citenamefont {Chin}, \citenamefont {Engeser}, \citenamefont
  {Lange}, \citenamefont {Pilch}, \citenamefont {Jaakkola}, \citenamefont
  {N{\"a}gerl} \emph {et~al.}}]{kraemer2006evidence}%
  \BibitemOpen
  \bibfield  {author} {\bibinfo {author} {\bibfnamefont {T.}~\bibnamefont
  {Kraemer}}, \bibinfo {author} {\bibfnamefont {M.}~\bibnamefont {Mark}},
  \bibinfo {author} {\bibfnamefont {P.}~\bibnamefont {Waldburger}}, \bibinfo
  {author} {\bibfnamefont {J.}~\bibnamefont {Danzl}}, \bibinfo {author}
  {\bibfnamefont {C.}~\bibnamefont {Chin}}, \bibinfo {author} {\bibfnamefont
  {B.}~\bibnamefont {Engeser}}, \bibinfo {author} {\bibfnamefont
  {A.}~\bibnamefont {Lange}}, \bibinfo {author} {\bibfnamefont
  {K.}~\bibnamefont {Pilch}}, \bibinfo {author} {\bibfnamefont
  {A.}~\bibnamefont {Jaakkola}}, \bibinfo {author} {\bibfnamefont {H.-C.}\
  \bibnamefont {N{\"a}gerl}},  \emph {et~al.},\ }\href {\doibase
  10.1038/nature04626} {\bibfield  {journal} {\bibinfo  {journal} {Nature}\
  }\textbf {\bibinfo {volume} {440}},\ \bibinfo {pages} {315} (\bibinfo {year}
  {2006})}\BibitemShut {NoStop}%
\bibitem [{\citenamefont {Tung}\ \emph {et~al.}(2014)\citenamefont {Tung},
  \citenamefont {Jim\'enez-Garc\'{\i}a}, \citenamefont {Johansen},
  \citenamefont {Parker},\ and\ \citenamefont
  {Chin}}]{TungJimenez-GarciaJohansenEtAl2014a}%
  \BibitemOpen
  \bibfield  {author} {\bibinfo {author} {\bibfnamefont {S.-K.}\ \bibnamefont
  {Tung}}, \bibinfo {author} {\bibfnamefont {K.}~\bibnamefont
  {Jim\'enez-Garc\'{\i}a}}, \bibinfo {author} {\bibfnamefont {J.}~\bibnamefont
  {Johansen}}, \bibinfo {author} {\bibfnamefont {C.~V.}\ \bibnamefont
  {Parker}}, \ and\ \bibinfo {author} {\bibfnamefont {C.}~\bibnamefont
  {Chin}},\ }\href {\doibase 10.1103/PhysRevLett.113.240402} {\bibfield
  {journal} {\bibinfo  {journal} {Phys. Rev. Lett.}\ }\textbf {\bibinfo
  {volume} {113}},\ \bibinfo {pages} {240402} (\bibinfo {year}
  {2014})}\BibitemShut {NoStop}%
\bibitem [{\citenamefont {Pires}\ \emph {et~al.}(2014)\citenamefont {Pires},
  \citenamefont {Ulmanis}, \citenamefont {H\"afner}, \citenamefont {Repp},
  \citenamefont {Arias}, \citenamefont {Kuhnle},\ and\ \citenamefont
  {Weidem\"uller}}]{PiresUlmanisHaefnerEtAl2014}%
  \BibitemOpen
  \bibfield  {author} {\bibinfo {author} {\bibfnamefont {R.}~\bibnamefont
  {Pires}}, \bibinfo {author} {\bibfnamefont {J.}~\bibnamefont {Ulmanis}},
  \bibinfo {author} {\bibfnamefont {S.}~\bibnamefont {H\"afner}}, \bibinfo
  {author} {\bibfnamefont {M.}~\bibnamefont {Repp}}, \bibinfo {author}
  {\bibfnamefont {A.}~\bibnamefont {Arias}}, \bibinfo {author} {\bibfnamefont
  {E.~D.}\ \bibnamefont {Kuhnle}}, \ and\ \bibinfo {author} {\bibfnamefont
  {M.}~\bibnamefont {Weidem\"uller}},\ }\href {\doibase
  10.1103/PhysRevLett.112.250404} {\bibfield  {journal} {\bibinfo  {journal}
  {Phys. Rev. Lett.}\ }\textbf {\bibinfo {volume} {112}},\ \bibinfo {pages}
  {250404} (\bibinfo {year} {2014})}\BibitemShut {NoStop}%
\bibitem [{\citenamefont {Pollack}\ \emph {et~al.}(2009)\citenamefont
  {Pollack}, \citenamefont {Dries},\ and\ \citenamefont
  {Hulet}}]{pollack2009universality}%
  \BibitemOpen
  \bibfield  {author} {\bibinfo {author} {\bibfnamefont {S.~E.}\ \bibnamefont
  {Pollack}}, \bibinfo {author} {\bibfnamefont {D.}~\bibnamefont {Dries}}, \
  and\ \bibinfo {author} {\bibfnamefont {R.~G.}\ \bibnamefont {Hulet}},\ }\href
  {\doibase 10.1126/science.1182840} {\bibfield  {journal} {\bibinfo  {journal}
  {Science}\ }\textbf {\bibinfo {volume} {326}},\ \bibinfo {pages} {1683}
  (\bibinfo {year} {2009})}\BibitemShut {NoStop}%
\bibitem [{\citenamefont {Gross}\ \emph {et~al.}(2009)\citenamefont {Gross},
  \citenamefont {Shotan}, \citenamefont {Kokkelmans},\ and\ \citenamefont
  {Khaykovich}}]{GrossShotanKokkelmansEtAl2009}%
  \BibitemOpen
  \bibfield  {author} {\bibinfo {author} {\bibfnamefont {N.}~\bibnamefont
  {Gross}}, \bibinfo {author} {\bibfnamefont {Z.}~\bibnamefont {Shotan}},
  \bibinfo {author} {\bibfnamefont {S.}~\bibnamefont {Kokkelmans}}, \ and\
  \bibinfo {author} {\bibfnamefont {L.}~\bibnamefont {Khaykovich}},\ }\href
  {\doibase 10.1103/PhysRevLett.103.163202} {\bibfield  {journal} {\bibinfo
  {journal} {Phys. Rev. Lett.}\ }\textbf {\bibinfo {volume} {103}},\ \bibinfo
  {pages} {163202} (\bibinfo {year} {2009})}\BibitemShut {NoStop}%
\bibitem [{\citenamefont {Lompe}\ \emph {et~al.}(2010)\citenamefont {Lompe},
  \citenamefont {Ottenstein}, \citenamefont {Serwane}, \citenamefont {Wenz},
  \citenamefont {Z{\"u}rn},\ and\ \citenamefont
  {Jochim}}]{LompeOttensteinSerwaneEtAl2010}%
  \BibitemOpen
  \bibfield  {author} {\bibinfo {author} {\bibfnamefont {T.}~\bibnamefont
  {Lompe}}, \bibinfo {author} {\bibfnamefont {T.~B.}\ \bibnamefont
  {Ottenstein}}, \bibinfo {author} {\bibfnamefont {F.}~\bibnamefont {Serwane}},
  \bibinfo {author} {\bibfnamefont {A.~N.}\ \bibnamefont {Wenz}}, \bibinfo
  {author} {\bibfnamefont {G.}~\bibnamefont {Z{\"u}rn}}, \ and\ \bibinfo
  {author} {\bibfnamefont {S.}~\bibnamefont {Jochim}},\ }\href {\doibase
  10.1126/science.1193148} {\bibfield  {journal} {\bibinfo  {journal}
  {Science}\ }\textbf {\bibinfo {volume} {330}},\ \bibinfo {pages} {940}
  (\bibinfo {year} {2010})}\BibitemShut {NoStop}%
\bibitem [{\citenamefont {Nakajima}\ \emph {et~al.}(2011)\citenamefont
  {Nakajima}, \citenamefont {Horikoshi}, \citenamefont {Mukaiyama},
  \citenamefont {Naidon},\ and\ \citenamefont
  {Ueda}}]{NakajimaHorikoshiMukaiyamaEtAl2011}%
  \BibitemOpen
  \bibfield  {author} {\bibinfo {author} {\bibfnamefont {S.}~\bibnamefont
  {Nakajima}}, \bibinfo {author} {\bibfnamefont {M.}~\bibnamefont {Horikoshi}},
  \bibinfo {author} {\bibfnamefont {T.}~\bibnamefont {Mukaiyama}}, \bibinfo
  {author} {\bibfnamefont {P.}~\bibnamefont {Naidon}}, \ and\ \bibinfo {author}
  {\bibfnamefont {M.}~\bibnamefont {Ueda}},\ }\href {\doibase
  10.1103/PhysRevLett.106.143201} {\bibfield  {journal} {\bibinfo  {journal}
  {Phys. Rev. Lett.}\ }\textbf {\bibinfo {volume} {106}},\ \bibinfo {pages}
  {143201} (\bibinfo {year} {2011})}\BibitemShut {NoStop}%
\bibitem [{\citenamefont {Kunitski}\ \emph {et~al.}(2015)\citenamefont
  {Kunitski}, \citenamefont {Zeller}, \citenamefont {Voigtsberger},
  \citenamefont {Kalinin}, \citenamefont {Schmidt}, \citenamefont
  {Sch{\"o}ffler}, \citenamefont {Czasch}, \citenamefont {Sch{\"o}llkopf},
  \citenamefont {Grisenti}, \citenamefont {Jahnke}, \citenamefont {Blume},\
  and\ \citenamefont {D{\"o}rner}}]{KunitskiZellerVoigtsbergerEtAl2015}%
  \BibitemOpen
  \bibfield  {author} {\bibinfo {author} {\bibfnamefont {M.}~\bibnamefont
  {Kunitski}}, \bibinfo {author} {\bibfnamefont {S.}~\bibnamefont {Zeller}},
  \bibinfo {author} {\bibfnamefont {J.}~\bibnamefont {Voigtsberger}}, \bibinfo
  {author} {\bibfnamefont {A.}~\bibnamefont {Kalinin}}, \bibinfo {author}
  {\bibfnamefont {L.~P.~H.}\ \bibnamefont {Schmidt}}, \bibinfo {author}
  {\bibfnamefont {M.}~\bibnamefont {Sch{\"o}ffler}}, \bibinfo {author}
  {\bibfnamefont {A.}~\bibnamefont {Czasch}}, \bibinfo {author} {\bibfnamefont
  {W.}~\bibnamefont {Sch{\"o}llkopf}}, \bibinfo {author} {\bibfnamefont
  {R.~E.}\ \bibnamefont {Grisenti}}, \bibinfo {author} {\bibfnamefont
  {T.}~\bibnamefont {Jahnke}}, \bibinfo {author} {\bibfnamefont
  {D.}~\bibnamefont {Blume}}, \ and\ \bibinfo {author} {\bibfnamefont
  {R.}~\bibnamefont {D{\"o}rner}},\ }\href {\doibase 10.1126/science.aaa5601}
  {\bibfield  {journal} {\bibinfo  {journal} {Science}\ }\textbf {\bibinfo
  {volume} {348}},\ \bibinfo {pages} {551} (\bibinfo {year}
  {2015})}\BibitemShut {NoStop}%
\bibitem [{\citenamefont {Huang}\ \emph {et~al.}(2014)\citenamefont {Huang},
  \citenamefont {Sidorenkov}, \citenamefont {Grimm},\ and\ \citenamefont
  {Hutson}}]{huang2014observation}%
  \BibitemOpen
  \bibfield  {author} {\bibinfo {author} {\bibfnamefont {B.}~\bibnamefont
  {Huang}}, \bibinfo {author} {\bibfnamefont {L.~A.}\ \bibnamefont
  {Sidorenkov}}, \bibinfo {author} {\bibfnamefont {R.}~\bibnamefont {Grimm}}, \
  and\ \bibinfo {author} {\bibfnamefont {J.~M.}\ \bibnamefont {Hutson}},\
  }\href {\doibase 10.1103/PhysRevLett.112.190401} {\bibfield  {journal}
  {\bibinfo  {journal} {Phys. Rev. Lett.}\ }\textbf {\bibinfo {volume} {112}},\
  \bibinfo {pages} {190401} (\bibinfo {year} {2014})}\BibitemShut {NoStop}%
\bibitem [{\citenamefont {Miransky}(1980)}]{MIRANSKY1980421}%
  \BibitemOpen
  \bibfield  {author} {\bibinfo {author} {\bibfnamefont {V.}~\bibnamefont
  {Miransky}},\ }\href {\doibase https://doi.org/10.1016/0370-2693(80)91011-4}
  {\bibfield  {journal} {\bibinfo  {journal} {Physics Letters B}\ }\textbf
  {\bibinfo {volume} {91}},\ \bibinfo {pages} {421 } (\bibinfo {year}
  {1980})}\BibitemShut {NoStop}%
\bibitem [{\citenamefont {Pereira}\ \emph {et~al.}(2007)\citenamefont
  {Pereira}, \citenamefont {Nilsson},\ and\ \citenamefont
  {Castro~Neto}}]{PereiraNilssonCastroNeto2007}%
  \BibitemOpen
  \bibfield  {author} {\bibinfo {author} {\bibfnamefont {V.~M.}\ \bibnamefont
  {Pereira}}, \bibinfo {author} {\bibfnamefont {J.}~\bibnamefont {Nilsson}}, \
  and\ \bibinfo {author} {\bibfnamefont {A.~H.}\ \bibnamefont {Castro~Neto}},\
  }\href {\doibase 10.1103/PhysRevLett.99.166802} {\bibfield  {journal}
  {\bibinfo  {journal} {Phys. Rev. Lett.}\ }\textbf {\bibinfo {volume} {99}},\
  \bibinfo {pages} {166802} (\bibinfo {year} {2007})}\BibitemShut {NoStop}%
\bibitem [{\citenamefont {Shytov}\ \emph
  {et~al.}(2007{\natexlab{a}})\citenamefont {Shytov}, \citenamefont
  {Katsnelson},\ and\ \citenamefont {Levitov}}]{ShytovKatsnelsonLevitov2007}%
  \BibitemOpen
  \bibfield  {author} {\bibinfo {author} {\bibfnamefont {A.~V.}\ \bibnamefont
  {Shytov}}, \bibinfo {author} {\bibfnamefont {M.~I.}\ \bibnamefont
  {Katsnelson}}, \ and\ \bibinfo {author} {\bibfnamefont {L.~S.}\ \bibnamefont
  {Levitov}},\ }\href {\doibase 10.1103/PhysRevLett.99.236801} {\bibfield
  {journal} {\bibinfo  {journal} {Phys. Rev. Lett.}\ }\textbf {\bibinfo
  {volume} {99}},\ \bibinfo {pages} {236801} (\bibinfo {year}
  {2007}{\natexlab{a}})}\BibitemShut {NoStop}%
\bibitem [{\citenamefont {Shytov}\ \emph
  {et~al.}(2007{\natexlab{b}})\citenamefont {Shytov}, \citenamefont
  {Katsnelson},\ and\ \citenamefont {Levitov}}]{ShytovKatsnelsonLevitov2007d}%
  \BibitemOpen
  \bibfield  {author} {\bibinfo {author} {\bibfnamefont {A.~V.}\ \bibnamefont
  {Shytov}}, \bibinfo {author} {\bibfnamefont {M.~I.}\ \bibnamefont
  {Katsnelson}}, \ and\ \bibinfo {author} {\bibfnamefont {L.~S.}\ \bibnamefont
  {Levitov}},\ }\href {\doibase 10.1103/PhysRevLett.99.246802} {\bibfield
  {journal} {\bibinfo  {journal} {Phys. Rev. Lett.}\ }\textbf {\bibinfo
  {volume} {99}},\ \bibinfo {pages} {246802} (\bibinfo {year}
  {2007}{\natexlab{b}})}\BibitemShut {NoStop}%
\bibitem [{\citenamefont {Dong}(2011)}]{dong2011wave}%
  \BibitemOpen
  \bibfield  {author} {\bibinfo {author} {\bibfnamefont {S.-H.}\ \bibnamefont
  {Dong}},\ }\href@noop {} {\emph {\bibinfo {title} {Wave Equations in Higher
  Dimensions}}}\ (\bibinfo  {publisher} {Springer},\ \bibinfo {year}
  {2011})\BibitemShut {NoStop}%
\bibitem [{\citenamefont {Friedrich}(2013)}]{Friedrich2013b}%
  \BibitemOpen
  \bibfield  {author} {\bibinfo {author} {\bibfnamefont {H.}~\bibnamefont
  {Friedrich}},\ }\href {\doibase 10.1007/978-3-662-48526-2} {\enquote
  {\bibinfo {title} {Scattering theory},}\ } (\bibinfo {year}
  {2013})\BibitemShut {NoStop}%
\bibitem [{\citenamefont {Yang}(1987)}]{Yang1987b}%
  \BibitemOpen
  \bibfield  {author} {\bibinfo {author} {\bibfnamefont {C.~N.}\ \bibnamefont
  {Yang}},\ }\href {http://projecteuclid.org/euclid.cmp/1104159815} {\bibfield
  {journal} {\bibinfo  {journal} {Comm. Math. Phys.}\ }\textbf {\bibinfo
  {volume} {112}},\ \bibinfo {pages} {205} (\bibinfo {year}
  {1987})}\BibitemShut {NoStop}%
\bibitem [{\citenamefont {Ovdat}\ \emph {et~al.}(2017)\citenamefont {Ovdat},
  \citenamefont {Mao}, \citenamefont {Jiang}, \citenamefont {Andrei},\ and\
  \citenamefont {Akkermans}}]{OvdatMaoJiangEtAl2017}%
  \BibitemOpen
  \bibfield  {author} {\bibinfo {author} {\bibfnamefont {O.}~\bibnamefont
  {Ovdat}}, \bibinfo {author} {\bibfnamefont {J.}~\bibnamefont {Mao}}, \bibinfo
  {author} {\bibfnamefont {Y.}~\bibnamefont {Jiang}}, \bibinfo {author}
  {\bibfnamefont {E.~Y.}\ \bibnamefont {Andrei}}, \ and\ \bibinfo {author}
  {\bibfnamefont {E.}~\bibnamefont {Akkermans}},\ }\href
  {https://doi.org/10.1038/s41467-017-00591-8} {\bibfield  {journal} {\bibinfo
  {journal} {Nature Communications}\ }\textbf {\bibinfo {volume} {8}},\
  \bibinfo {pages} {507} (\bibinfo {year} {2017})}\BibitemShut {NoStop}%
\bibitem [{\citenamefont {Pereira}\ \emph {et~al.}(2008)\citenamefont
  {Pereira}, \citenamefont {Kotov},\ and\ \citenamefont
  {Castro~Neto}}]{PereiraKotovCastroNeto2008a}%
  \BibitemOpen
  \bibfield  {author} {\bibinfo {author} {\bibfnamefont {V.~M.}\ \bibnamefont
  {Pereira}}, \bibinfo {author} {\bibfnamefont {V.~N.}\ \bibnamefont {Kotov}},
  \ and\ \bibinfo {author} {\bibfnamefont {A.~H.}\ \bibnamefont
  {Castro~Neto}},\ }\href {\doibase 10.1103/PhysRevB.78.085101} {\bibfield
  {journal} {\bibinfo  {journal} {Phys. Rev. B}\ }\textbf {\bibinfo {volume}
  {78}},\ \bibinfo {pages} {085101} (\bibinfo {year} {2008})}\BibitemShut
  {NoStop}%
\bibitem [{\citenamefont {Ovdat}\ \emph {et~al.}(2018)\citenamefont {Ovdat},
  \citenamefont {Don},\ and\ \citenamefont {Akkermans}}]{ovdat2018vacancies}%
  \BibitemOpen
  \bibfield  {author} {\bibinfo {author} {\bibfnamefont {O.}~\bibnamefont
  {Ovdat}}, \bibinfo {author} {\bibfnamefont {Y.}~\bibnamefont {Don}}, \ and\
  \bibinfo {author} {\bibfnamefont {E.}~\bibnamefont {Akkermans}},\ }\href
  {https://arxiv.org/pdf/1807.10297.pdf} {\bibfield  {journal} {\bibinfo
  {journal} {arXiv preprint arXiv:1807.10297}\ } (\bibinfo {year}
  {2018})}\BibitemShut {NoStop}%
\bibitem [{\citenamefont {Katsnelson}(2012)}]{Katsnelson2012d}%
  \BibitemOpen
  \bibfield  {author} {\bibinfo {author} {\bibfnamefont {M.~I.}\ \bibnamefont
  {Katsnelson}},\ }\href@noop {} {\emph {\bibinfo {title} {Graphene: carbon in
  two dimensions}}}\ (\bibinfo  {publisher} {Cambridge University Press, New
  York},\ \bibinfo {year} {2012})\BibitemShut {NoStop}%
\bibitem [{\citenamefont {Katsnelson}\ \emph {et~al.}(2006)\citenamefont
  {Katsnelson}, \citenamefont {Novoselov},\ and\ \citenamefont
  {Geim}}]{KatsnelsonNovoselovGeim2006}%
  \BibitemOpen
  \bibfield  {author} {\bibinfo {author} {\bibfnamefont {M.~I.}\ \bibnamefont
  {Katsnelson}}, \bibinfo {author} {\bibfnamefont {K.~S.}\ \bibnamefont
  {Novoselov}}, \ and\ \bibinfo {author} {\bibfnamefont {A.~K.}\ \bibnamefont
  {Geim}},\ }\href {http://dx.doi.org/10.1038/nphys384} {\bibfield  {journal}
  {\bibinfo  {journal} {Nat Phys}\ }\textbf {\bibinfo {volume} {2}},\ \bibinfo
  {pages} {620} (\bibinfo {year} {2006})}\BibitemShut {NoStop}%
\bibitem [{\citenamefont {Stander}\ \emph {et~al.}(2009)\citenamefont
  {Stander}, \citenamefont {Huard},\ and\ \citenamefont
  {Goldhaber-Gordon}}]{PhysRevLett.102.026807}%
  \BibitemOpen
  \bibfield  {author} {\bibinfo {author} {\bibfnamefont {N.}~\bibnamefont
  {Stander}}, \bibinfo {author} {\bibfnamefont {B.}~\bibnamefont {Huard}}, \
  and\ \bibinfo {author} {\bibfnamefont {D.}~\bibnamefont {Goldhaber-Gordon}},\
  }\href {\doibase 10.1103/PhysRevLett.102.026807} {\bibfield  {journal}
  {\bibinfo  {journal} {Phys. Rev. Lett.}\ }\textbf {\bibinfo {volume} {102}},\
  \bibinfo {pages} {026807} (\bibinfo {year} {2009})}\BibitemShut {NoStop}%
\bibitem [{\citenamefont {Zhang}\ \emph {et~al.}(2005)\citenamefont {Zhang},
  \citenamefont {Tan}, \citenamefont {Stormer},\ and\ \citenamefont
  {Kim}}]{ZhangTanStormerEtAl2005}%
  \BibitemOpen
  \bibfield  {author} {\bibinfo {author} {\bibfnamefont {Y.}~\bibnamefont
  {Zhang}}, \bibinfo {author} {\bibfnamefont {Y.-W.}\ \bibnamefont {Tan}},
  \bibinfo {author} {\bibfnamefont {H.~L.}\ \bibnamefont {Stormer}}, \ and\
  \bibinfo {author} {\bibfnamefont {P.}~\bibnamefont {Kim}},\ }\href
  {http://dx.doi.org/10.1038/nature04235} {\bibfield  {journal} {\bibinfo
  {journal} {Nature}\ }\textbf {\bibinfo {volume} {438}},\ \bibinfo {pages}
  {201} (\bibinfo {year} {2005})}\BibitemShut {NoStop}%
\bibitem [{\citenamefont {Wang}\ \emph {et~al.}(2013)\citenamefont {Wang},
  \citenamefont {Wong}, \citenamefont {Shytov}, \citenamefont {Brar},
  \citenamefont {Choi}, \citenamefont {Wu}, \citenamefont {Tsai}, \citenamefont
  {Regan}, \citenamefont {Zettl}, \citenamefont {Kawakami}, \citenamefont
  {Louie}, \citenamefont {Levitov},\ and\ \citenamefont
  {Crommie}}]{WangWongShytovEtAl2013a}%
  \BibitemOpen
  \bibfield  {author} {\bibinfo {author} {\bibfnamefont {Y.}~\bibnamefont
  {Wang}}, \bibinfo {author} {\bibfnamefont {D.}~\bibnamefont {Wong}}, \bibinfo
  {author} {\bibfnamefont {A.~V.}\ \bibnamefont {Shytov}}, \bibinfo {author}
  {\bibfnamefont {V.~W.}\ \bibnamefont {Brar}}, \bibinfo {author}
  {\bibfnamefont {S.}~\bibnamefont {Choi}}, \bibinfo {author} {\bibfnamefont
  {Q.}~\bibnamefont {Wu}}, \bibinfo {author} {\bibfnamefont {H.-Z.}\
  \bibnamefont {Tsai}}, \bibinfo {author} {\bibfnamefont {W.}~\bibnamefont
  {Regan}}, \bibinfo {author} {\bibfnamefont {A.}~\bibnamefont {Zettl}},
  \bibinfo {author} {\bibfnamefont {R.~K.}\ \bibnamefont {Kawakami}}, \bibinfo
  {author} {\bibfnamefont {S.~G.}\ \bibnamefont {Louie}}, \bibinfo {author}
  {\bibfnamefont {L.~S.}\ \bibnamefont {Levitov}}, \ and\ \bibinfo {author}
  {\bibfnamefont {M.~F.}\ \bibnamefont {Crommie}},\ }\href {\doibase
  10.1126/science.1234320} {\bibfield  {journal} {\bibinfo  {journal}
  {Science}\ }\textbf {\bibinfo {volume} {340}},\ \bibinfo {pages} {734}
  (\bibinfo {year} {2013})}\BibitemShut {NoStop}%
\bibitem [{\citenamefont {Mao}\ \emph {et~al.}(2016)\citenamefont {Mao},
  \citenamefont {Jiang}, \citenamefont {Moldovan}, \citenamefont {Li},
  \citenamefont {Watanabe}, \citenamefont {Taniguchi}, \citenamefont {Masir},
  \citenamefont {Peeters},\ and\ \citenamefont {Andrei}}]{mao2016realization}%
  \BibitemOpen
  \bibfield  {author} {\bibinfo {author} {\bibfnamefont {J.}~\bibnamefont
  {Mao}}, \bibinfo {author} {\bibfnamefont {Y.}~\bibnamefont {Jiang}}, \bibinfo
  {author} {\bibfnamefont {D.}~\bibnamefont {Moldovan}}, \bibinfo {author}
  {\bibfnamefont {G.}~\bibnamefont {Li}}, \bibinfo {author} {\bibfnamefont
  {K.}~\bibnamefont {Watanabe}}, \bibinfo {author} {\bibfnamefont
  {T.}~\bibnamefont {Taniguchi}}, \bibinfo {author} {\bibfnamefont {M.~R.}\
  \bibnamefont {Masir}}, \bibinfo {author} {\bibfnamefont {F.~M.}\ \bibnamefont
  {Peeters}}, \ and\ \bibinfo {author} {\bibfnamefont {E.~Y.}\ \bibnamefont
  {Andrei}},\ }\href {http://dx.doi.org/10.1038/nphys3665} {\bibfield
  {journal} {\bibinfo  {journal} {Nat. Phys.}\ }\textbf {\bibinfo {volume}
  {12}},\ \bibinfo {pages} {545} (\bibinfo {year} {2016})}\BibitemShut
  {NoStop}%
\bibitem [{\citenamefont {Liu}\ \emph {et~al.}(2015)\citenamefont {Liu},
  \citenamefont {Weinert},\ and\ \citenamefont {Li}}]{LiuWeinertLi2015}%
  \BibitemOpen
  \bibfield  {author} {\bibinfo {author} {\bibfnamefont {Y.}~\bibnamefont
  {Liu}}, \bibinfo {author} {\bibfnamefont {M.}~\bibnamefont {Weinert}}, \ and\
  \bibinfo {author} {\bibfnamefont {L.}~\bibnamefont {Li}},\ }\href
  {http://stacks.iop.org/0957-4484/26/i=3/a=035702} {\bibfield  {journal}
  {\bibinfo  {journal} {Nanotechnology}\ }\textbf {\bibinfo {volume} {26}},\
  \bibinfo {pages} {035702} (\bibinfo {year} {2015})}\BibitemShut {NoStop}%
\bibitem [{\citenamefont {Akkermans}\ and\ \citenamefont
  {Montambaux}(2007)}]{akkermans2007mesoscopic}%
  \BibitemOpen
  \bibfield  {author} {\bibinfo {author} {\bibfnamefont {E.}~\bibnamefont
  {Akkermans}}\ and\ \bibinfo {author} {\bibfnamefont {G.}~\bibnamefont
  {Montambaux}},\ }in\ \href {\doibase
  https://doi.org/10.1017/CBO9780511618833} {\emph {\bibinfo {booktitle}
  {{Mesoscopic Physics of Electrons and Photons}}}}\ (\bibinfo  {publisher}
  {Cambridge University Press},\ \bibinfo {year} {2007})\ Chap.~\bibinfo
  {chapter} {7}\BibitemShut {NoStop}%
\bibitem [{\citenamefont {Gorsky}\ and\ \citenamefont
  {Popov}(2014)}]{gorsky2014atomic}%
  \BibitemOpen
  \bibfield  {author} {\bibinfo {author} {\bibfnamefont {A.}~\bibnamefont
  {Gorsky}}\ and\ \bibinfo {author} {\bibfnamefont {F.}~\bibnamefont {Popov}},\
  }\href {\doibase 10.1103/PhysRevD.89.061702} {\bibfield  {journal} {\bibinfo
  {journal} {Phys. Rev. D}\ }\textbf {\bibinfo {volume} {89}},\ \bibinfo
  {pages} {061702} (\bibinfo {year} {2014})}\BibitemShut {NoStop}%
\bibitem [{\citenamefont {Brattan}\ \emph
  {et~al.}(2018{\natexlab{a}})\citenamefont {Brattan}, \citenamefont {Ovdat},\
  and\ \citenamefont {Akkermans}}]{BrattanPhysRevD.97.061701}%
  \BibitemOpen
  \bibfield  {author} {\bibinfo {author} {\bibfnamefont {D.~K.}\ \bibnamefont
  {Brattan}}, \bibinfo {author} {\bibfnamefont {O.}~\bibnamefont {Ovdat}}, \
  and\ \bibinfo {author} {\bibfnamefont {E.}~\bibnamefont {Akkermans}},\ }\href
  {\doibase 10.1103/PhysRevD.97.061701} {\bibfield  {journal} {\bibinfo
  {journal} {Phys. Rev. D}\ }\textbf {\bibinfo {volume} {97}},\ \bibinfo
  {pages} {061701} (\bibinfo {year} {2018}{\natexlab{a}})}\BibitemShut
  {NoStop}%
\bibitem [{\citenamefont {Alexandre}(2011)}]{Alexandre2011}%
  \BibitemOpen
  \bibfield  {author} {\bibinfo {author} {\bibfnamefont {J.}~\bibnamefont
  {Alexandre}},\ }\href {\doibase 10.1142/S0217751X11054656} {\bibfield
  {journal} {\bibinfo  {journal} {Int. J. Mod. Phys.}\ }\textbf {\bibinfo
  {volume} {A26}},\ \bibinfo {pages} {4523} (\bibinfo {year}
  {2011})}\BibitemShut {NoStop}%
\bibitem [{\citenamefont {Hornreich}\ \emph {et~al.}(1975)\citenamefont
  {Hornreich}, \citenamefont {Luban},\ and\ \citenamefont
  {Shtrikman}}]{PhysRevLett.35.1678}%
  \BibitemOpen
  \bibfield  {author} {\bibinfo {author} {\bibfnamefont {R.~M.}\ \bibnamefont
  {Hornreich}}, \bibinfo {author} {\bibfnamefont {M.}~\bibnamefont {Luban}}, \
  and\ \bibinfo {author} {\bibfnamefont {S.}~\bibnamefont {Shtrikman}},\ }\href
  {\doibase 10.1103/PhysRevLett.35.1678} {\bibfield  {journal} {\bibinfo
  {journal} {Phys. Rev. Lett.}\ }\textbf {\bibinfo {volume} {35}},\ \bibinfo
  {pages} {1678} (\bibinfo {year} {1975})}\BibitemShut {NoStop}%
\bibitem [{\citenamefont {Grinstein}(1981)}]{PhysRevB.23.4615}%
  \BibitemOpen
  \bibfield  {author} {\bibinfo {author} {\bibfnamefont {G.}~\bibnamefont
  {Grinstein}},\ }\href {\doibase 10.1103/PhysRevB.23.4615} {\bibfield
  {journal} {\bibinfo  {journal} {Phys. Rev. B}\ }\textbf {\bibinfo {volume}
  {23}},\ \bibinfo {pages} {4615} (\bibinfo {year} {1981})}\BibitemShut
  {NoStop}%
\bibitem [{\citenamefont {Fradkin}\ \emph {et~al.}(2004)\citenamefont
  {Fradkin}, \citenamefont {Huse}, \citenamefont {Moessner}, \citenamefont
  {Oganesyan},\ and\ \citenamefont {Sondhi}}]{FradkinHuseMoessnerEtAl2004}%
  \BibitemOpen
  \bibfield  {author} {\bibinfo {author} {\bibfnamefont {E.}~\bibnamefont
  {Fradkin}}, \bibinfo {author} {\bibfnamefont {D.~A.}\ \bibnamefont {Huse}},
  \bibinfo {author} {\bibfnamefont {R.}~\bibnamefont {Moessner}}, \bibinfo
  {author} {\bibfnamefont {V.}~\bibnamefont {Oganesyan}}, \ and\ \bibinfo
  {author} {\bibfnamefont {S.~L.}\ \bibnamefont {Sondhi}},\ }\href {\doibase
  10.1103/PhysRevB.69.224415} {\bibfield  {journal} {\bibinfo  {journal} {Phys.
  Rev. B}\ }\textbf {\bibinfo {volume} {69}},\ \bibinfo {pages} {224415}
  (\bibinfo {year} {2004})}\BibitemShut {NoStop}%
\bibitem [{\citenamefont {Vishwanath}\ \emph {et~al.}(2004)\citenamefont
  {Vishwanath}, \citenamefont {Balents},\ and\ \citenamefont
  {Senthil}}]{VishwanathBalentsSenthil2004}%
  \BibitemOpen
  \bibfield  {author} {\bibinfo {author} {\bibfnamefont {A.}~\bibnamefont
  {Vishwanath}}, \bibinfo {author} {\bibfnamefont {L.}~\bibnamefont {Balents}},
  \ and\ \bibinfo {author} {\bibfnamefont {T.}~\bibnamefont {Senthil}},\ }\href
  {\doibase 10.1103/PhysRevB.69.224416} {\bibfield  {journal} {\bibinfo
  {journal} {Phys. Rev. B}\ }\textbf {\bibinfo {volume} {69}},\ \bibinfo
  {pages} {224416} (\bibinfo {year} {2004})}\BibitemShut {NoStop}%
\bibitem [{\citenamefont {Ardonne}\ \emph {et~al.}(2004)\citenamefont
  {Ardonne}, \citenamefont {Fendley},\ and\ \citenamefont
  {Fradkin}}]{ArdonneFendleyFradkin2004}%
  \BibitemOpen
  \bibfield  {author} {\bibinfo {author} {\bibfnamefont {E.}~\bibnamefont
  {Ardonne}}, \bibinfo {author} {\bibfnamefont {P.}~\bibnamefont {Fendley}}, \
  and\ \bibinfo {author} {\bibfnamefont {E.}~\bibnamefont {Fradkin}},\ }\href
  {\doibase 10.1016/j.aop.2004.01.004} {\bibfield  {journal} {\bibinfo
  {journal} {Annals Phys.}\ }\textbf {\bibinfo {volume} {310}},\ \bibinfo
  {pages} {493} (\bibinfo {year} {2004})}\BibitemShut {NoStop}%
\bibitem [{\citenamefont {McCann}\ and\ \citenamefont
  {Koshino}(2013)}]{0034-4885-76-5-056503}%
  \BibitemOpen
  \bibfield  {author} {\bibinfo {author} {\bibfnamefont {E.}~\bibnamefont
  {McCann}}\ and\ \bibinfo {author} {\bibfnamefont {M.}~\bibnamefont
  {Koshino}},\ }\href {http://stacks.iop.org/0034-4885/76/i=5/a=056503}
  {\bibfield  {journal} {\bibinfo  {journal} {Reports on Progress in Physics}\
  }\textbf {\bibinfo {volume} {76}},\ \bibinfo {pages} {056503} (\bibinfo
  {year} {2013})}\BibitemShut {NoStop}%
\bibitem [{\citenamefont {{Ramires}}\ \emph {et~al.}(2012)\citenamefont
  {{Ramires}}, \citenamefont {{Coleman}}, \citenamefont {{Nevidomskyy}},\ and\
  \citenamefont {{Tsvelik}}}]{2012PhRvL.109q6404R}%
  \BibitemOpen
  \bibfield  {author} {\bibinfo {author} {\bibfnamefont {A.}~\bibnamefont
  {{Ramires}}}, \bibinfo {author} {\bibfnamefont {P.}~\bibnamefont
  {{Coleman}}}, \bibinfo {author} {\bibfnamefont {A.~H.}\ \bibnamefont
  {{Nevidomskyy}}}, \ and\ \bibinfo {author} {\bibfnamefont {A.~M.}\
  \bibnamefont {{Tsvelik}}},\ }\href {\doibase 10.1103/PhysRevLett.109.176404}
  {\bibfield  {journal} {\bibinfo  {journal} {Physical Review Letters}\
  }\textbf {\bibinfo {volume} {109}},\ \bibinfo {eid} {176404} (\bibinfo {year}
  {2012})},\ \Eprint {http://arxiv.org/abs/1207.6441} {arXiv:1207.6441
  [cond-mat.str-el]} \BibitemShut {NoStop}%
\bibitem [{\citenamefont {Mukohyama}(2010)}]{Mukohyama2010}%
  \BibitemOpen
  \bibfield  {author} {\bibinfo {author} {\bibfnamefont {S.}~\bibnamefont
  {Mukohyama}},\ }\href {\doibase 10.1088/0264-9381/27/22/223101} {\bibfield
  {journal} {\bibinfo  {journal} {Class. Quant. Grav.}\ }\textbf {\bibinfo
  {volume} {27}},\ \bibinfo {pages} {223101} (\bibinfo {year}
  {2010})}\BibitemShut {NoStop}%
\bibitem [{\citenamefont {Kachru}\ \emph {et~al.}(2008)\citenamefont {Kachru},
  \citenamefont {Liu},\ and\ \citenamefont {Mulligan}}]{KachruLiuMulligan2008}%
  \BibitemOpen
  \bibfield  {author} {\bibinfo {author} {\bibfnamefont {S.}~\bibnamefont
  {Kachru}}, \bibinfo {author} {\bibfnamefont {X.}~\bibnamefont {Liu}}, \ and\
  \bibinfo {author} {\bibfnamefont {M.}~\bibnamefont {Mulligan}},\ }\href
  {\doibase 10.1103/PhysRevD.78.106005} {\bibfield  {journal} {\bibinfo
  {journal} {Phys. Rev.}\ }\textbf {\bibinfo {volume} {D78}},\ \bibinfo {pages}
  {106005} (\bibinfo {year} {2008})}\BibitemShut {NoStop}%
\bibitem [{\citenamefont {Horava}(2009{\natexlab{a}})}]{Horava2009}%
  \BibitemOpen
  \bibfield  {author} {\bibinfo {author} {\bibfnamefont {P.}~\bibnamefont
  {Horava}},\ }\href {\doibase 10.1103/PhysRevD.79.084008} {\bibfield
  {journal} {\bibinfo  {journal} {Phys. Rev.}\ }\textbf {\bibinfo {volume}
  {D79}},\ \bibinfo {pages} {084008} (\bibinfo {year}
  {2009}{\natexlab{a}})}\BibitemShut {NoStop}%
\bibitem [{\citenamefont {Horava}(2009{\natexlab{b}})}]{Horava2009a}%
  \BibitemOpen
  \bibfield  {author} {\bibinfo {author} {\bibfnamefont {P.}~\bibnamefont
  {Horava}},\ }\href {\doibase 10.1103/PhysRevLett.102.161301} {\bibfield
  {journal} {\bibinfo  {journal} {Phys. Rev. Lett.}\ }\textbf {\bibinfo
  {volume} {102}},\ \bibinfo {pages} {161301} (\bibinfo {year}
  {2009}{\natexlab{b}})}\BibitemShut {NoStop}%
\bibitem [{\citenamefont {Strogatz}(2014)}]{strogatz2014nonlinear}%
  \BibitemOpen
  \bibfield  {author} {\bibinfo {author} {\bibfnamefont {S.~H.}\ \bibnamefont
  {Strogatz}},\ }\href@noop {} {\emph {\bibinfo {title} {Nonlinear dynamics and
  chaos: with applications to physics, biology, chemistry, and engineering}}}\
  (\bibinfo  {publisher} {Westview press},\ \bibinfo {year} {2014})\BibitemShut
  {NoStop}%
\bibitem [{\citenamefont {Brattan}\ \emph
  {et~al.}(2018{\natexlab{b}})\citenamefont {Brattan}, \citenamefont {Ovdat},\
  and\ \citenamefont {Akkermans}}]{OvdatJounalOfPhysA}%
  \BibitemOpen
  \bibfield  {author} {\bibinfo {author} {\bibfnamefont {D.~K.}\ \bibnamefont
  {Brattan}}, \bibinfo {author} {\bibfnamefont {O.}~\bibnamefont {Ovdat}}, \
  and\ \bibinfo {author} {\bibfnamefont {E.}~\bibnamefont {Akkermans}},\ }\href
  {http://stacks.iop.org/1751-8121/51/i=43/a=435401} {\bibfield  {journal}
  {\bibinfo  {journal} {Journal of Physics A: Mathematical and Theoretical}\
  }\textbf {\bibinfo {volume} {51}},\ \bibinfo {pages} {435401} (\bibinfo
  {year} {2018}{\natexlab{b}})}\BibitemShut {NoStop}%
\bibitem [{\citenamefont {Appelquist}\ \emph {et~al.}(1988)\citenamefont
  {Appelquist}, \citenamefont {Nash},\ and\ \citenamefont
  {Wijewardhana}}]{AppelquistNashWijewardhana1988}%
  \BibitemOpen
  \bibfield  {author} {\bibinfo {author} {\bibfnamefont {T.}~\bibnamefont
  {Appelquist}}, \bibinfo {author} {\bibfnamefont {D.}~\bibnamefont {Nash}}, \
  and\ \bibinfo {author} {\bibfnamefont {L.~C.~R.}\ \bibnamefont
  {Wijewardhana}},\ }\href {\doibase 10.1103/PhysRevLett.60.2575} {\bibfield
  {journal} {\bibinfo  {journal} {Phys. Rev. Lett.}\ }\textbf {\bibinfo
  {volume} {60}},\ \bibinfo {pages} {2575} (\bibinfo {year}
  {1988})}\BibitemShut {NoStop}%
\bibitem [{\citenamefont {Herbut}(2016)}]{HerbutPhysRevD.94.025036}%
  \BibitemOpen
  \bibfield  {author} {\bibinfo {author} {\bibfnamefont {I.~F.}\ \bibnamefont
  {Herbut}},\ }\href {\doibase 10.1103/PhysRevD.94.025036} {\bibfield
  {journal} {\bibinfo  {journal} {Phys. Rev. D}\ }\textbf {\bibinfo {volume}
  {94}},\ \bibinfo {pages} {025036} (\bibinfo {year} {2016})}\BibitemShut
  {NoStop}%
\bibitem [{\citenamefont {Nielsen}\ \emph {et~al.}(2001)\citenamefont
  {Nielsen}, \citenamefont {Fedorov}, \citenamefont {Jensen},\ and\
  \citenamefont {Garrido}}]{NIELSEN2001373}%
  \BibitemOpen
  \bibfield  {author} {\bibinfo {author} {\bibfnamefont {E.}~\bibnamefont
  {Nielsen}}, \bibinfo {author} {\bibfnamefont {D.}~\bibnamefont {Fedorov}},
  \bibinfo {author} {\bibfnamefont {A.}~\bibnamefont {Jensen}}, \ and\ \bibinfo
  {author} {\bibfnamefont {E.}~\bibnamefont {Garrido}},\ }\href {\doibase
  https://doi.org/10.1016/S0370-1573(00)00107-1} {\bibfield  {journal}
  {\bibinfo  {journal} {Physics Reports}\ }\textbf {\bibinfo {volume} {347}},\
  \bibinfo {pages} {373 } (\bibinfo {year} {2001})}\BibitemShut {NoStop}%
\bibitem [{\citenamefont {Mohapatra}\ and\ \citenamefont
  {Braaten}(2018)}]{PhysRevA.98.013633}%
  \BibitemOpen
  \bibfield  {author} {\bibinfo {author} {\bibfnamefont {A.}~\bibnamefont
  {Mohapatra}}\ and\ \bibinfo {author} {\bibfnamefont {E.}~\bibnamefont
  {Braaten}},\ }\href {\doibase 10.1103/PhysRevA.98.013633} {\bibfield
  {journal} {\bibinfo  {journal} {Phys. Rev. A}\ }\textbf {\bibinfo {volume}
  {98}},\ \bibinfo {pages} {013633} (\bibinfo {year} {2018})}\BibitemShut
  {NoStop}%
\bibitem [{\citenamefont {Bedaque}\ \emph {et~al.}(1999)\citenamefont
  {Bedaque}, \citenamefont {Hammer},\ and\ \citenamefont {van
  Kolck}}]{BedaqueHammerVanKloPhysRevLett.82.463}%
  \BibitemOpen
  \bibfield  {author} {\bibinfo {author} {\bibfnamefont {P.~F.}\ \bibnamefont
  {Bedaque}}, \bibinfo {author} {\bibfnamefont {H.-W.}\ \bibnamefont {Hammer}},
  \ and\ \bibinfo {author} {\bibfnamefont {U.}~\bibnamefont {van Kolck}},\
  }\href {\doibase 10.1103/PhysRevLett.82.463} {\bibfield  {journal} {\bibinfo
  {journal} {Phys. Rev. Lett.}\ }\textbf {\bibinfo {volume} {82}},\ \bibinfo
  {pages} {463} (\bibinfo {year} {1999})}\BibitemShut {NoStop}%
\bibitem [{\citenamefont {Deng}\ \emph {et~al.}(2016)\citenamefont {Deng},
  \citenamefont {Shi}, \citenamefont {Diao}, \citenamefont {Yu}, \citenamefont
  {Zhai}, \citenamefont {Qi},\ and\ \citenamefont {Wu}}]{Deng2016}%
  \BibitemOpen
  \bibfield  {author} {\bibinfo {author} {\bibfnamefont {S.}~\bibnamefont
  {Deng}}, \bibinfo {author} {\bibfnamefont {Z.-Y.}\ \bibnamefont {Shi}},
  \bibinfo {author} {\bibfnamefont {P.}~\bibnamefont {Diao}}, \bibinfo {author}
  {\bibfnamefont {Q.}~\bibnamefont {Yu}}, \bibinfo {author} {\bibfnamefont
  {H.}~\bibnamefont {Zhai}}, \bibinfo {author} {\bibfnamefont {R.}~\bibnamefont
  {Qi}}, \ and\ \bibinfo {author} {\bibfnamefont {H.}~\bibnamefont {Wu}},\
  }\href {\doibase 10.1126/science.aaf0666} {\bibfield  {journal} {\bibinfo
  {journal} {Science}\ }\textbf {\bibinfo {volume} {353}},\ \bibinfo {pages}
  {371} (\bibinfo {year} {2016})}\BibitemShut {NoStop}%
\end{thebibliography}%

\end{document}